\documentclass[
    prl, twocolumn, superscriptaddress, nofootinbib, amsmath, amssymb,
    aps, floatfix, preprintnumbers
]{revtex4-2}
\usepackage[utf8]{inputenc}
\usepackage{setspace}
\usepackage[dvips]{graphicx}
\usepackage[dvipsnames]{xcolor}
\definecolor{CiteBlue}{RGB}{45,52,151}
\usepackage[
    colorlinks=true,
    linkcolor=CiteBlue,
    urlcolor=CiteBlue,
    citecolor=CiteBlue
]{hyperref}
\usepackage{aas_macros}
\usepackage{tikz}

\usepackage[capitalise]{cleveref}
\usepackage{siunitx}
\DeclareSIUnit{\year}{yr}

\let\oldsection\section

\newcommand{\refcite}[1]{Ref.~\cite{#1}}
\newcommand{\refscite}[1]{Refs.~\cite{#1}}

\usepackage{bm}
\newcommand{\bb}[1]{\bm{\mathrm{#1}}}
\newcommand{\du}{\mathrm d}

\renewcommand{\Im}{\operatorname{Im}}
\newcommand{\dm}{\chi}
\newcommand{\med}{\phi}
\newcommand{\qp}{\mathrm{QP}}

\newcommand{\plas}{\mathrm{p}}

\begin{document}
\title{Detecting Light Dark Matter with Kinetic Inductance Detectors}
\preprint{MIT-CTP/5654}

\author{Jiansong Gao}
\thanks{Currently at Amazon Web Services Center for Quantum Computing, Pasadena, CA 91006.}
\affiliation{National Institute of Standards and Technology, Boulder, CO 80305, USA}

\author{Yonit Hochberg}
\affiliation{Racah Institute of Physics, Hebrew University of Jerusalem, Jerusalem 91904, Israel}

\author{Benjamin V. Lehmann}
\affiliation{Center for Theoretical Physics, Massachusetts Institute of Technology, Cambridge, MA 02139, USA}

\author{Sae Woo Nam}
\thanks{Deceased.}
\affiliation{National Institute of Standards and Technology, Boulder, CO 80305, USA}

\author{Paul Szypryt}
\affiliation{National Institute of Standards and Technology, Boulder, CO 80305, USA}
\affiliation{Department of Physics, University of Colorado, Boulder, CO 80309, USA}

\author{Michael R. Vissers}
\affiliation{National Institute of Standards and Technology, Boulder, CO 80305, USA}

\author{Tao Xu}
\affiliation{Homer L. Dodge Department of Physics and Astronomy, University of Oklahoma, Norman, OK 73019, USA}

\date\today

\begin{abstract}\ignorespaces
    Superconducting detectors are a promising technology for probing dark matter at extremely low masses, where dark matter interactions are currently unconstrained. Realizing the potential of such detectors requires new readout technologies to achieve the lowest possible thresholds for deposited energy. Here we perform a prototype search for dark matter--electron interactions with kinetic inductance detectors (KIDs), a class of superconducting detector originally designed for infrared astronomy applications. We demonstrate that existing KIDs can achieve effective thresholds as low as \qty{0.2}{\electronvolt}, and we use existing data to set new dark matter constraints. The relative maturity of the technology underlying KIDs means that this platform can be scaled significantly with existing tools, enabling powerful new searches in the coming years.
\end{abstract}

\maketitle

The nature of dark matter (DM) remains one of the most significant open problems in particle physics and cosmology. In the wake of decades of null searches for DM particles at the weak scale, the focus of the community has shifted to particles in different mass ranges, especially at sub-GeV masses, where traditional DM detection experiments lose sensitivity. At these masses, the detection of elastic DM--nucleon scattering faces a severe challenge due to unfavorable kinematics: the mismatch between the mass of the DM and that of the nuclear target limits the energy deposited in the scattering process. On the other hand, the detection of DM--electron interactions is much more readily viable, due to both the lower mass of the electron and the energy scales associated with electronic excitations in detector systems.

The key challenge in designing such experiments is to achieve sensitivity to extremely small amounts of deposited energy. Numerous experimental concepts have been proposed for this purpose, including detectors based on semiconductors \cite{Essig:2015cda,Graham:2012su,Hochberg:2016sqx,Griffin:2020lgd}, superconductors \cite{Hochberg:2015pha,Hochberg:2019cyy,Hochberg:2021yud}, two-dimensional targets~\cite{Hochberg:2016ntt,Cavoto:2017otc}, superfluid Helium \cite{Schutz:2016tid,Knapen:2016cue}, and Dirac materials~\cite{Hochberg:2017wce}. In each case, the threshold of the detector is tied to the lowest-lying excitations of these gapped systems. In particular, superconductors exhibit gaps below $\sim$\qty{1}{\milli\electronvolt}, meaning that they can in principle probe DM--electron scattering at DM masses below $\sim$\qty{1}{\kilo\electronvolt}. Below this scale, cosmological observations severely restrict fermionic DM, making this an important target for the next generation of experiments.

Superconductors have already been employed to set competitive limits on sub-MeV DM in experiments based on superconducting nanowire single-photon detectors~(SNSPDs) \cite{Hochberg:2019cyy,Hochberg:2021yud}. However, realizing the potential of superconducting detectors to probe the lowest DM masses still requires substantial technological development. The small gap of a superconductor arises from the binding energy of Cooper pairs: a small amount of deposited energy can break a pair and produce a pair of quasiparticle excitations. Detecting the presence of these quasiparticle excitations is challenging. 
Present-day experiments perform the readout based on the macroscopic effects of quasiparticle production. In particular, SNSPDs rely on heating: significant quasiparticle production leads to a large number of secondary phonons, causing a sharp transition from the superconducting phase to the normal metal phase. 
However, such detectors have limited energy resolution: their readout method does not distinguish between different deposit sizes as long as they are sufficient to trigger a transition.
Additionally, constructing a large-scale DM detector with such small sensors requires massive multiplexing, which is challenging for SNSPDs. While new readout techniques are beginning to make large SNSPD arrays feasible \cite{Oripov:2023yod}, substantial research and development will be required for future DM searches. This motivates the parallel development of other readout methodologies.

In this work, we prototype a new type of superconducting DM experiment aimed at resolving these issues. For the first time, we perform a low-threshold search for DM--electron interactions using a kinetic inductance detector (KID) \cite{mazin2002superconducting,Day:2003,Zmuidzinas:2012,Gao:2012rb,Baselmans:2012,Guo:2017ukv,Mazin:2019xkb}. KIDs have previously been used to search for DM--nucleon scattering in \refscite{Golwala:2008,Cruciani:2022mbb,Pagnanini:2015ulo}. KIDs take advantage of the fact that the kinetic inductance of a superconductor increases when charge-carrying Cooper pairs are broken. When patterned into a LC resonator, shifts in the kinetic inductance can be precisely measured with standard microwave electronics, enabling a very sensitive readout without the sharp transition-based threshold of SNSPDs. Small-scale KIDs have already been developed for use in astronomical applications in the optical and near-infrared regimes \cite{2013PASP..125.1348M, 2018PASP..130f5001M, 2020PASP..132l5005W, OBrien:2020uu}. The simplicity and relative maturity of the technology means that it is feasible to produce devices at large scales with present-day readout schemes \cite{McHugh:2012dy,Fruitwala:2020ibn,CCAT-prime:2022tho}, so KIDs have the potential to significantly extend DM searches in both mass and cross section. Even in this prototype study, we demonstrate excellent energy resolution with thresholds as low as \qty{0.2}{\electronvolt}, substantially outperforming other searches to date.

\section{Kinetic inductance detectors}
\label{sec:kids}
The operation of a KID is based on the inverse relationship between the kinetic inductance of a superconductor and the density of charge carriers. Energy deposits in the superconductor break Cooper pairs, producing excess quasiparticles and increasing its kinetic inductance. When energy is deposited in the superconductor, the resonance frequency $f(t)$ decreases rapidly with the quasiparticle density. The quasiparticles recombine on a timescale $\tau_\qp$, after which $f(t)$ relaxes back to the quiescent resonance frequency $\bar f$. We read out the changing resonance frequency using a homodyne mixing scheme, and search for events in the frequency quadrature.\footnote{For further details, see Section 5.2 of \refcite{GaoThesis:2008}.} We determine the signal shape in the frequency quadrature empirically from calibration data, and we use this shape to identify energy deposition events and reject noise.

The sensitivity of KIDs derives from the high kinetic inductance fraction of superconducting strips in materials with high resistivity in the normal metal state. Typical KIDs operate in the microwave range,\footnote{For this reason, these devices are sometimes referred to in the literature as microwave kinetic inductance detectors (MKIDs).} with resonance frequencies $\bar f\sim\qty{}{\giga\hertz}$, and the noise level in our readout system is $\delta \bar f\sim\qty{}{\kilo\hertz}$, enabling the detection of shifts in the resonance frequency at one part in \num{e6}. For sufficiently high-quality resonators, which are readily fabricated with standard techniques, the magnitude of the frequency shift follows a Gaussian distribution whose mean and variance depends on the energy. For photons, neglecting any noise sources in the readout, the maximum energy resolving power $R\equiv E/\Delta E_\mathrm{FWHM}$ has the analytic form
\begin{equation}
    \label{eq:Rmax}
    R_{\mathrm{max}} =
    \frac{1}{2.355}\sqrt{\frac{\eta E}{F\Delta}}
    ,
\end{equation}
where $\eta$ is the conversion efficiency from photons to quasiparticles, $\Delta \approx 1.76k_\mathrm{B}T_{C}$ is the superconducting gap, $F$ is the Fano factor, and the factor of $1/2.355$ converts between the standard deviation of the energy distribution and the full width at half maximum (FWHM). Thus, in particular, a small superconducting gap corresponds to a high energy resolving power.

For our experimental configuration, we have $\eta \approx 0.57$, $T_C \approx \qty{1.2}{\kelvin}$, $\Delta \approx \qty{0.18}{\milli\electronvolt}$, and $F \approx 0.2$, corresponding to $R_{\mathrm{max}} = 53.1\times(E/\qty{}{\electronvolt})^{1/2}$. This readily enables the detection of sub-eV photons, which is why KIDs have been deployed for infrared astronomy applications. For our purposes, we note that $R_{\mathrm{max}} \gtrsim 5$ for energies as low as \qty{10}{\milli\electronvolt}, meaning that this detection platform can in principle retain sensitivity and energy resolution for deposits characteristic of DM masses as low as \qty{10}{\kilo\electronvolt}. Crucially, the readout scheme allows such resonators to be massively multiplexed, enabling straightforward implementation of experiments consisting of large arrays of KIDs. KIDs can also be easily operated cryogenically at temperatures well below the superconducting transition temperature, which allows for thorough suppression of noise from thermal quasiparticles.

\section{Implementation and calibration}
\label{sec:experiment}

\begin{figure}\centering
    \includegraphics{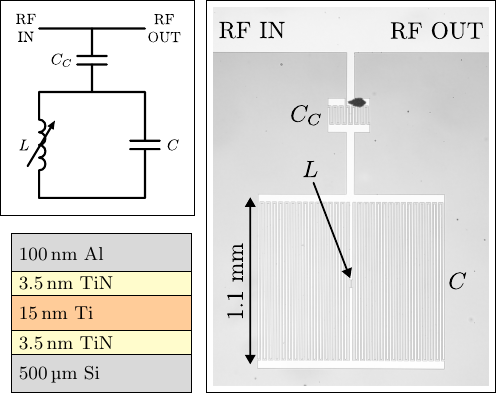}
    \caption{\textbf{Top left:} Schematic representation of a KID and readout as a circuit. \textbf{Bottom left:} Layer structure including substrate and Al cap. \textbf{Right:} Micrograph of the KID used in this study, with circuit elements labeled.}
    \label{fig:detector}
\end{figure}

We constructed a kinetic inductance detector as the sensor and target for a prototype DM search. The kinetic inductance detector used in this study, shown in \cref{fig:detector}, is similar to that used in \refcite{Guo:2017ukv}. The superconducting absorber, i.e., the active component of the detector, consists of three layers: one of Ti between two of TiN, with a critical temperature of $T_C \approx \qty{1.2}{\kelvin}$. The absorber has a total thickness of \qty{22}{\nano\meter}, and a surface area of $\qty{1}{\micro\meter} \times \qty{50}{\micro\meter}$. The absorber is placed in series with an interdigitated capacitor (IDC) and an inductive strip in order to form an RLC resonator. The IDC has a much larger area than the inductor on the order of $\sim\!\!\qty{1}{\milli\meter^2}$, and is capped with a thin Al layer, which attenuates variations in the effective inductance of the resonator due to the IDC. It is possible that excitations produced by energy deposition in the IDC could be transported to the absorber, breaking Cooper pairs and producing a signal. However, as the efficiency for this process is difficult to estimate as a function of location, we conservatively neglect all deposits in the IDC when evaluating DM sensitivity curves. We assume the internal efficiency to energy deposition is $100\%$, which is confirmed by comparing the single photon detection rate and the photon absorption rate of the device studied in \refcite{Guo:2017ukv}. The KID's response to energy deposition can be measured in the frequency and the dissipation responses. We use frequency readout in this study, as is typical in sensing applications: the frequency response has a larger amplitude and longer lifetime, and thus yields a much higher signal-to-noise ratio.

Counting events requires a prescription for identifying events in timeseries data. Low-threshold count detectors such as SNSPDs count events based on sharp transitions between the superconducting and normal-metal phases, which are pronounced and easily identified. For a KID, by contrast, a sufficiently small deposit can shift the resonance frequency by an almost arbitrarily small amount.\footnote{The absolute floor is fixed by the typical value of $\omega$ in quasiparticle production processes, which is set by the superconducting gap, well below any scale considered in this work.} Discriminating between events and noise fluctuations is nontrivial, and the details of the analysis procedure have important implications for sensitivity to the smallest deposits.

\begin{figure}\centering
    \includegraphics[width=\columnwidth]{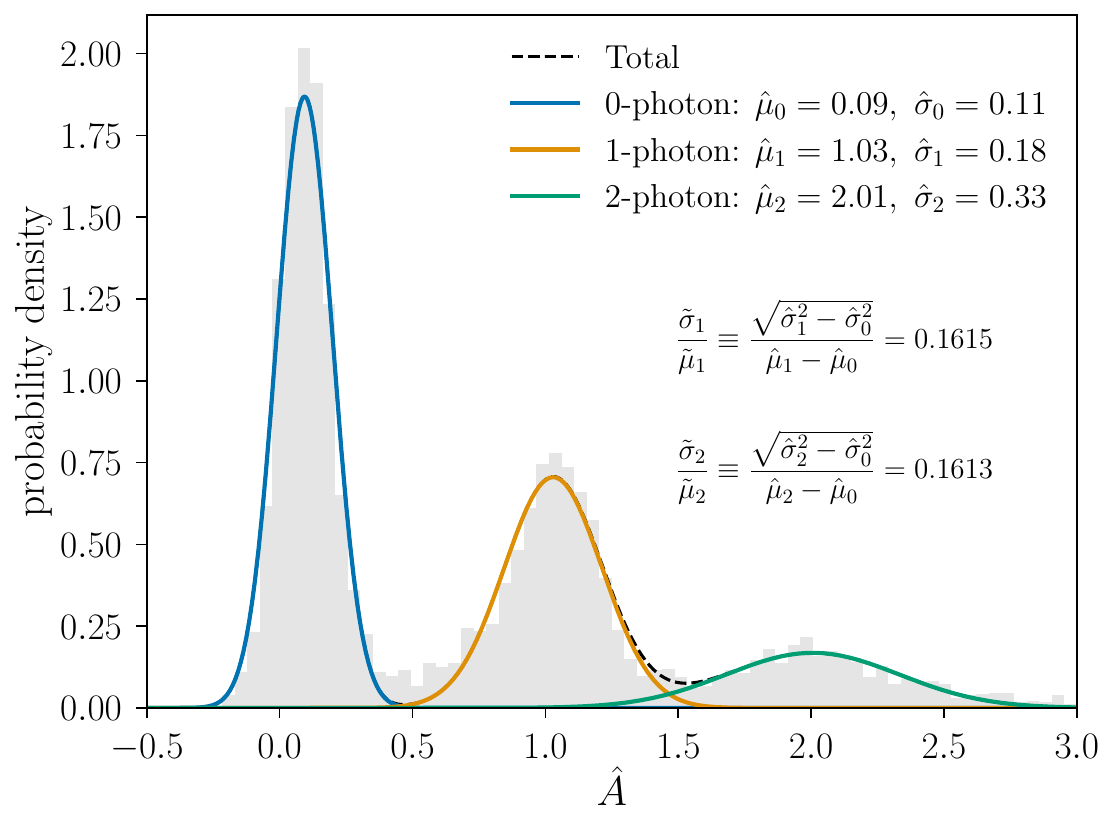}
    \caption{Distribution of estimated pulse heights $\hat A$ on calibration data with a \qty{0.8}{\electronvolt} photon source. Blue, orange, and green curves show Gaussian fits to the 0-photon, 1-photon, and 2-photon peaks, respectively. Note that $\tilde\sigma_n/\tilde\mu_n$ is nearly constant across $n$-photon peaks.}
    \label{fig:n-photon-peaks}
\end{figure}

We identify candidate events by matching timeseries samples of the frequency quadrature, $\Delta f(t)$, to a template pulse shape $s(t)$, which we extract empirically from a calibration dataset. We obtained the calibration data by exposing the detector to a \qty{0.8}{\electronvolt} (\qty{1550}{\nano\meter}) laser pulse at regular intervals. Given a measured timeseries $\Delta f(t)$ of the same duration $T$ as the template $s(t)$, we write $\Delta f(t) = As(t) + n(t)$, where $n(t)$ accounts for noise. We then extract $A$ from the data using a standard Wiener filter, obtaining an estimate $\hat A$. Using a sliding window of duration $T$ gives a timeseries $\hat A(t)$. A deposit of energy $\omega$ will produce a peak in $\hat A(t)$ with height corresponding to $\omega$. Further details are given in the Supplementary Material (SM).

In particular, this means that the effects of energy deposits on KID data can be phrased in terms of the \textit{distribution} of the estimator $\hat A$ over a set of samples, as shown for the calibration data itself in \cref{fig:n-photon-peaks}. The peak of the distribution at $\hat A \approx 1$ corresponds to the absorption of a single photon from the calibration source, and the peak at $\hat A \approx 2$ corresponds to occasional two-photon absorption events. The width of the ``zero-photon'' peak at small $\hat A$ characterizes the distribution of $\hat A$ in the absence of a calibration photon, which reflects the intrinsic noise of the detector. This width $\hat\sigma_0$ is ultimately the limiting factor on the sensitivity to small deposits. Note that the energy resolving power of our detector at the calibration energy is $R \approx 2.4$, well below the upper limit of $R_{\mathrm{max}} = 44.2$ implied by \cref{eq:Rmax}.

\Cref{fig:n-photon-peaks} demonstrates the simple relation between $\hat A$ and the size of the deposit, $\omega$. We normalize $\hat A$ to 1 for the \qty{0.8}{\electronvolt} calibration pulses. Then the distance between the $n$-photon peak and the 0-photon peak, $\tilde\mu_n\equiv\hat\mu_n - \hat\mu_0$, is linear in $\omega$. We also extract an empirical relationship between the peak location and its width. For fixed detector parameters, there is no mechanism intrinsic to KIDs that causes the resolution to degrade with increasing photon energy. However, it is clear from \cref{fig:n-photon-peaks} that $\hat\sigma_n$ increases with $\hat\mu_n$. The same effect was observed in \refcite{Guo:2017ukv} and attributed to additional noise induced by the absorption of the photon. Taking this photon-induced noise $\tilde\sigma_n$ to combine in quadrature with the width of the 0-photon peak, we define $\tilde\sigma_n \equiv \sqrt{\hat\sigma_n^2 - \hat\sigma_0^2}$, where $\hat\sigma_n$ is the width of the $n$-photon peak. Empirically, $\tilde\sigma_n / \tilde\mu_n$ is nearly constant at 0.16, suggesting that the additional noise is linear in the deposited energy. This allows us to predict the noise induced by deposited energies that are not multiples of the calibration energy. In particular, we map between $\hat A$ and $\omega$ using $(\hat\mu_0, \hat\sigma_0) = (0.09, 0.11)$. Note that $\hat\mu_0 > 0$ in the calibration dataset, i.e., the 0-photon peak is slightly displaced from $\hat A = 0$. This is due at least in part to the absorption of a large number of calibration photons in the IDC: the IDC has a large area, but the response of the detector to deposits therein is very small.

\begin{figure}\centering
    \includegraphics[width=\columnwidth]{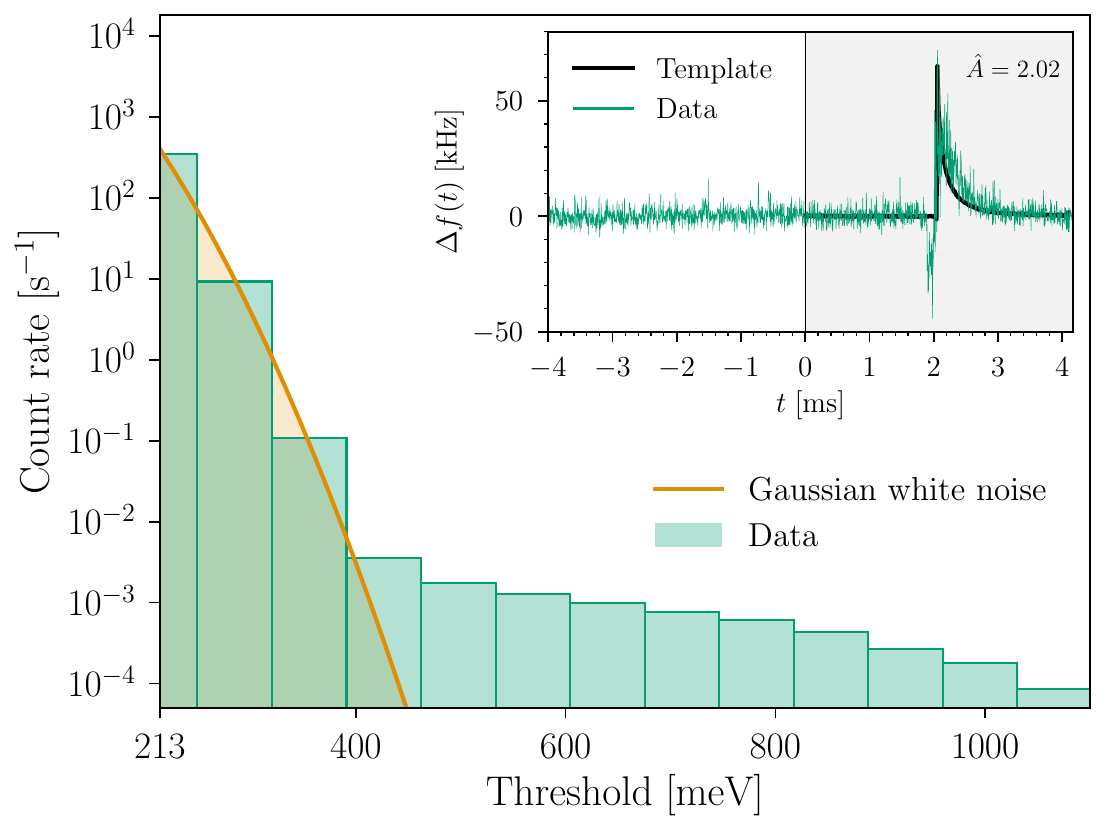}
    \caption{Integrated count rate observed in the prototype dataset as a function of effective threshold. The orange curve shows the dark count rate predicted in the case of pure Gaussian white noise with variance set by the measured 0-photon peak (see \cref{fig:n-photon-peaks}). \textbf{Inset:} The single event in the dataset with the largest value of $\hat A$ (2.02). The timeseries measurements $\Delta f(t)$ are shown in green, while the scaled signal template $\hat A\times s(t)$ is shown in black in the shaded region. The time is given relative to the start of the signal template.}
    \label{fig:dark-count-rate}
\end{figure}

We operated the detector in dark conditions, with no calibration source, for 26 hours. We use the event rate from this run to constrain DM interactions in the following section. To that end, we must first define the counting of events. In a click detector, such as an SNSPD, the count rate is simply the number of clicks per unit time. For our present analysis, on the other hand, the count rate is a function of the threshold in $\hat A$, which corresponds directly to the threshold in the deposited energy. We thus obtain a spectrum of count rates as a function of $\hat A$, or equivalently, as a function of $\omega$, as shown in \cref{fig:dark-count-rate}. We then diagnose the presence or absence of a DM signal using the profile likelihood ratio test to distinguish between Gaussian white noise alone and the combination of noise and a DM signal.

Since we can reduce the threshold in $\hat A$ almost arbitrarily, it is possible to study DM interactions at very low energy thresholds, with the caveat that the dark count rate increases dramatically at small $\hat A$. Even these high dark count rates are not necessarily an obstacle to obtaining strong constraints if the noise distribution is sufficiently well-characterized. However, since the noise characterization is not completely controlled, we conservatively set a threshold of \qty{200}{\milli\electronvolt} for our analysis.

We note that while the observed dark count rate at low energies is consistent with Gaussian white noise associated with the width of the 0-photon peak, this is not the case for energies $\omega \gtrsim \qty{0.5}{\electronvolt}$. \Cref{fig:dark-count-rate} shows a long tail of dark counts that cannot be accounted for by white noise. These counts could result from an energetic source, such as cosmic rays or radioactive materials, or from non-Gaussian electronic noise. In our projections for future experiments, we will optimistically assume that the dark counts are fully accounted for by Gaussian white noise.

\section{Dark matter constraints}
\label{sec:dm-interactions}

\begin{figure*}\centering
    \includegraphics[width=0.49\textwidth]{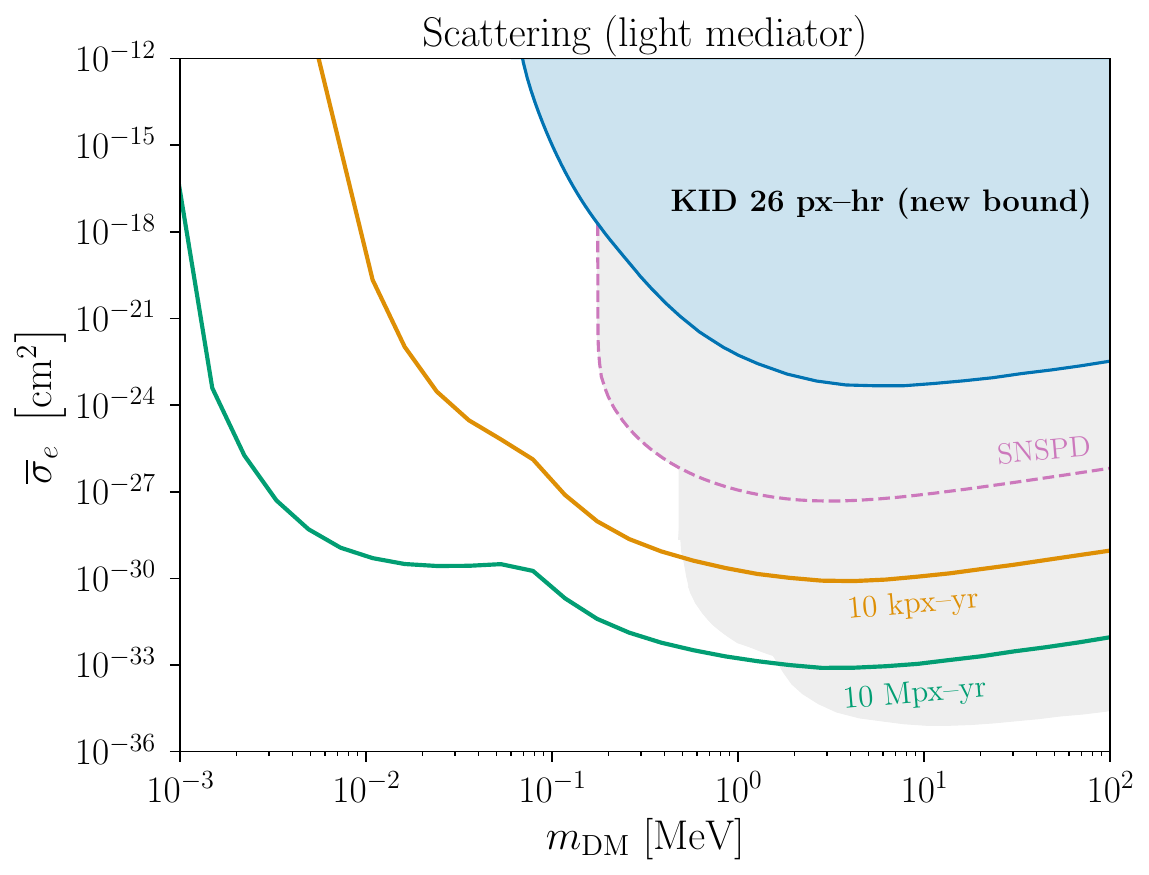}
    \hfill
    \includegraphics[width=\columnwidth]{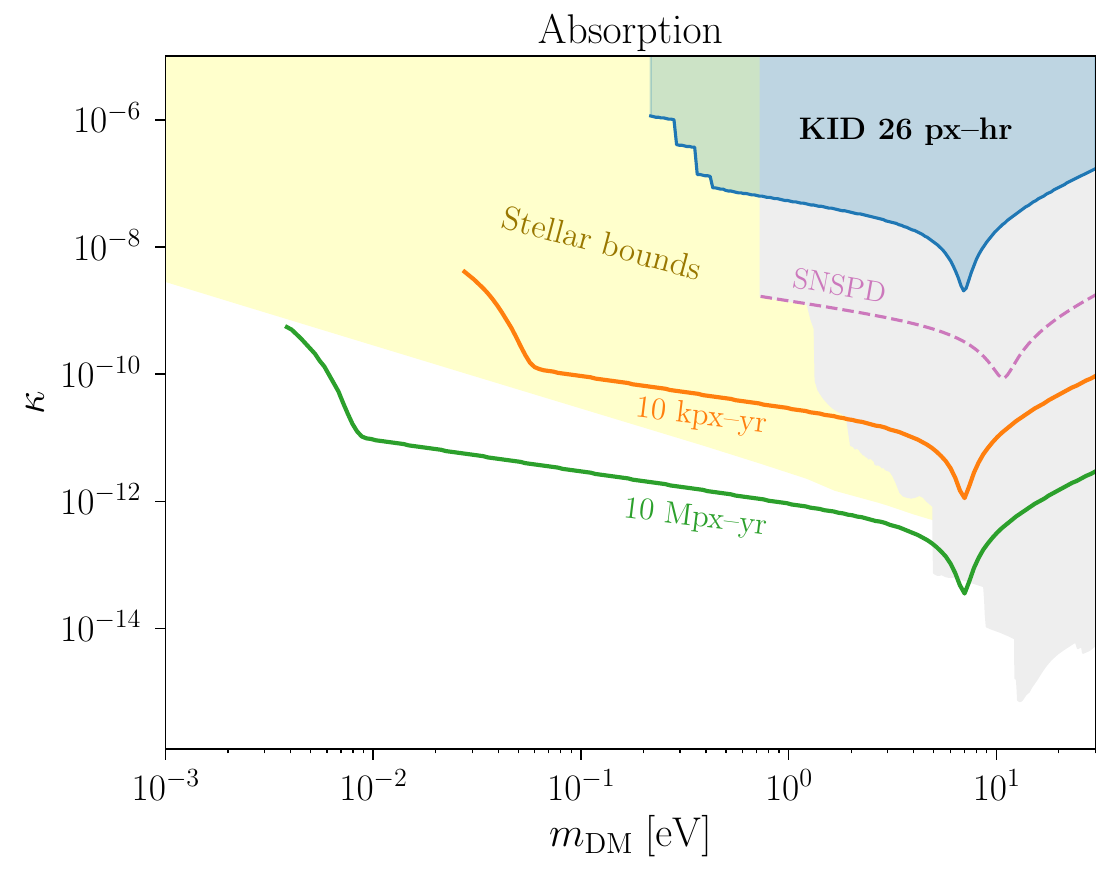}
    \caption{\textbf{Left:} New bound and projections for DM scattering with electrons via a light mediator. The shaded gray region shows existing bounds from terrestrial experiments \cite{Barak:2020fql,Amaral:2020ryn,Aguilar-Arevalo:2019wdi,Essig:2017kqs,Agnes:2018oej,Aprile:2019xxb}, with recent SNSPD results outlined in dashed purple. The orange and green curves show projected constraints for year-long exposures of multiplexed KIDs with \num{e4} and \num{e7} pixels, respectively, each of the same size as our single-pixel prototype. In the projected curves, we assume that the high-energy dark counts in \cref{fig:dark-count-rate} are mitigated, such that background counts are produced only by the Gaussian white noise associated with the 0-photon peak. The width of this peak is reduced for the projected curves, corresponding to an increased energy resolving power and a decreased threshold. The resolving power for the orange curve is increased by a factor of 8, matching demonstrated resolving power in comparable devices. The green curve assumes the maximum resolving power in a low-gap absorber material with $T_C = \qty{100}{\milli\kelvin}$ ($R=136.2$). \textbf{Right:} New bound and projections for absorption of kinetically mixed dark photon DM. The shaded gray region shows existing bounds from terrestrial experiments \cite{An:2014twa,Agnese:2018col,Aguilar-Arevalo:2019wdi,Arnaud:2020svb,FUNKExperiment:2020ofv,Barak:2020fql}, and the shaded yellow region shows model-dependent bounds from stellar cooling \cite{An:2013yua,An:2014twa,An:2020bxd}.}
    \label{fig:reach}
\end{figure*}

The interaction rate of DM with electrons in a detector is determined by three components: (1) the abundance and distribution of dark matter particles passing through the experiment; (2) the form of the microphysical interaction between the DM and a single electron; and (3) the electronic degrees of freedom in the detector material. The first two components admit a set of benchmark models that have been well understood for many years. The third component, relating to the materials physics itself, has only recently been explored in the literature, and has been shown to have significant implications for experimental sensitivity. We assume that the DM--electron interaction is spin-independent, which allows us to use the dielectric function formalism for DM scattering rates \cite{Hochberg:2021pkt}.

In this approach, the dependence of the interaction rate on the electronic degrees of freedom is written in terms of a linear response function $\chi(\bb q, \omega)$ for any momentum transfer $\bb q$ and energy deposit $\omega$. The DM scattering rate at fixed DM velocity $\bb v_\dm$ then takes the form
\begin{equation}
    \label{eq:scattering-rate}
    \Gamma(\bb v_\dm) = \int\frac{\du^3\bb q}{(2\pi)^3}|V(\bb q)|^2\left[
        2\chi(\bb q, \omega_{\bb q})
    \right]
    \,.
\end{equation}
Here $V(\bb q)\equiv g_eg_\chi/(\bb q^2+m_\med^2)$ is the DM--electron interaction potential, where $m_\med$ is the mediator mass, and $g_e$ and $g_\dm$ are the couplings of the mediator to the electron and the DM particle respectively; $\omega_{\bb q} = \bb q\cdot\bb v_\dm - \bb q^2/(2m_\dm)$; and $\chi(\bb q, \omega)$ is related directly to the dielectric function $\epsilon(\bb q, \omega)$ as $\chi(\bb q, \omega) = (\bb q^2/e^2)\Im[-1/\epsilon(\bb q, \omega)]$, where $e$ is the charge of the electron. The \textit{loss function} $\Im(-1/\epsilon)$ is a well understood and experimentally measurable property of an electronic system. The rate of a DM absorption process is also proportional to the loss function, as we review in the SM.

We use this form of the rate to determine constraints on the DM-electron cross section for the scattering case, and on the dark photon mixing parameter for the case of absorption. Our constraints for scattering are shown in terms of a reference cross section $\bar\sigma_e = \frac1\pi\mu_{e\dm}^2g_e^2g_\dm^2\bigl[
        \left(\alpha_{\mathrm{EM}}m_e\right)^2 + m_\med^2
    \bigr]^{-2}$, where $\mu_{e\dm}$ denotes the reduced mass of the electron--DM system, and $\alpha_{\mathrm{EM}} \approx 1/137$ is the fine structure constant.
We assume the DM velocities to be distributed according to the Standard Halo Model, taking the rms velocity as \qty{220}{\kilo\meter/\second}, the Earth velocity as \qty{232}{\kilo\meter/\second} in the galactic frame, and the halo escape velocity as \qty{550}{\kilo\meter/\second}. 

The resulting bounds we set using the collected data from our prototype KID are shown in \cref{fig:reach} for DM scattering via a light mediator ({\it left panel}) and for absorption of a kinetically mixed dark photon ({\it right panel}). Results for DM scattering via a heavy mediator are shown in the SM. We also show projections for the reach of future KID-based experiments, assuming improvements in both exposure and threshold. In particular, the orange curves in \cref{fig:reach} assume that the energy resolving power is increased by a uniform factor to $R = 19$ at \qty{0.8}{\electronvolt}, as has been demonstrated by \refcite{deVisser:2021kip}. This is a factor of $\sim\!8$ higher than $R$ in our configuration, but still well below $R_{\mathrm{max}} \approx 60$. The green curves assume maximum resolving power in a material with a smaller gap, corresponding to a critical temperature of $T_C = \qty{100}{\milli\kelvin}$, or $R_{\mathrm{max}} = 136.2$. A critical temperature of this order could be achieved in a future experiment using e.g. Hf or a proximitized bilayer for the absorber. The data from our prototype experiment already probes new parameter space for DM scattering due to the extremely low effective threshold. It also extends direct detection constraints for dark photon absorption into parameter space that is probed by stellar physics in a complementary manner.

\section{Discussion}
\label{sec:discussion}
In this work, we have demonstrated the use of KIDs as DM detectors as simultaneous targets and sensors, and we have shown that our prototype device is sensitive to lower thresholds than any existing DM experiment, including those based on SNSPDs. The remarkably low threshold of the KID experiment originates from the detection concept: rather than identifying events by sharp transitions out of the superconducting phase, the KID registers even very small deposits as correspondingly small fluctuations in its resonance frequency. Thus, the main hurdle to overcome is noise rejection, and we have shown here that simple filtering techniques are sufficient to achieve a threshold of \qty{0.2}{\electronvolt}.

Additionally, our prototype device demonstrates excellent energy resolution for candidate events, as compared with other DM detection platforms. In calibration data, the energy $\omega$ of an incident pulse can be resolved with a full width at half maximum of $\sim\!\!0.3\omega$. In future large-scale experiments, this spectral information can be used to discriminate between a putative DM signal and known backgrounds. Moreover, our current results correspond to an energy resolving power more than an order of magnitude below the upper limit of \cref{eq:Rmax}, indicating considerable room for improvement by reduction of noise sources. Significantly higher energy resolving power has already been demonstrated in other KID devices.

Despite the extremely small size of our prototype device, it has already set novel constraints on DM interactions due to the extremely low effective threshold. The most promising aspect of KID-based DM searches is their scalability. KIDs are easily manufactured in large multiplexed arrays, and it is possible with present-day tools to produce KID arrays with active volumes on the order of \qty{1}{\milli\meter^3}, larger than our current setup by a factor of $\sim\!\!\num{e8}$. This increased size will translate directly to substantially enhanced reach in the DM parameter space. Moreover, future detector development will also reduce the 0-photon noise width, leading to yet smaller effective thresholds.

\begin{acknowledgments}
\textbf{Acknowledgments.} We thank Noah Kurinsky for useful discussions and for comments on the manuscript. 
The work of Y.H. is supported by the Israel Science Foundation (grant No. 1818/22), by the Binational Science Foundation (grants No. 2018140 and No. 2022287) and by an ERC STG grant (``Light-Dark'', grant No. 101040019). 
The work of B.V.L. is supported by the MIT Pappalardo Fellowship.
The work of T.X. is supported by DOE grant NO. DE-SC0009956.
This project has received funding from the European Research Council (ERC) under the European Union’s Horizon Europe research and innovation programme (grant agreement No. 101040019).  Views and opinions expressed are however those of the author(s) only and do not necessarily reflect those of the European Union. The European Union cannot be held responsible for them.

\textbf{In Memoriam.} We regretfully acknowledge the untimely passing of Sae Woo Nam during the final stages of this work. We dedicate this article to his memory.
\end{acknowledgments}

\bibliography{references}


\onecolumngrid
\clearpage

\setcounter{page}{1}
\setcounter{equation}{0}
\setcounter{figure}{0}
\setcounter{table}{0}
\setcounter{section}{0}
\setcounter{subsection}{0}
\renewcommand{\theequation}{S.\arabic{equation}}
\renewcommand{\thefigure}{S\arabic{figure}}
\renewcommand{\thetable}{S\arabic{table}}
\renewcommand{\thesection}{\Roman{section}}
\renewcommand{\thesubsection}{\Alph{subsection}}
\newcommand{\ssection}[1]{
    \addtocounter{section}{1}
    \oldsection{\thesection.~~~#1}
    \addtocounter{section}{-1}
    \refstepcounter{section}
    \noindent\ignorespaces
}
\newcommand{\ssubsection}[1]{
    \addtocounter{subsection}{1}
    \subsection{\thesubsection.~~~#1}
    \addtocounter{subsection}{-1}
    \refstepcounter{subsection}
    \noindent\ignorespaces
}
\newcommand{\fakeaffil}[2]{$^{#1}$\textit{#2}\\}

\thispagestyle{empty}
\begin{center}
    \begin{spacing}{1.2}
        \textbf{\large 
            Supplemental Material:\\
            Detecting Light Dark Matter with Kinetic Inductance Detectors
        }
    \end{spacing}
    \par\smallskip
    Jiansong Gao,\textsuperscript{1, $*$}
    Yonit  Hochberg,\textsuperscript{2}
    Benjamin V. Lehmann,\textsuperscript{3}
    Sae\\Woo Nam,\textsuperscript{1, $\dagger$}
    Paul Szypryt,\textsuperscript{1, 4}
    Michael R. Vissers,\textsuperscript{1}
    and Tao Xu\textsuperscript{5}
    \par\smallskip
    {\small
        \fakeaffil{1}{National Institute of Standards and Technology, Boulder, CO 80305, USA}
        \fakeaffil{2}{Racah Institute of Physics, Hebrew University of Jerusalem, Jerusalem 91904, Israel}
        \fakeaffil{3}{Center for Theoretical Physics, Massachusetts Institute of Technology, Cambridge, MA 02139, USA}
        \fakeaffil{4}{Department of Physics, University of Colorado, Boulder, CO 80309, USA}
        \fakeaffil{5}{Homer L. Dodge Department of Physics and Astronomy,\\University of Oklahoma, Norman, OK 73019, USA}
        (Dated: \today)
    }

\end{center}
\par\smallskip

In this Supplemental Material, we provide additional details on the analysis procedure used in this work to extract constraints on dark matter (DM) interactions.

\ssection{Optimal filtering and event selection}
Given the raw timeseries data $\Delta f(t)$, we identify candidate events using an optimal filter given a template event shape. To that end, we perform the following steps:
\begin{enumerate}
    \item Determine the pulse shape template $s(t)$ with duration $T$;
    \item Determine the spectrum of noise $n(t)$;
    \item Assuming the signal takes the form $\Delta f(t) = As(t) + n(t)$, estimate $A$ on each window of duration $T$ using the maximum likelihood estimator $\hat A$ given in \cref{eq:A-estimator} below, producing a timeseries $\hat A(t)$;
    \item Define criteria for counting the number of events from $\hat A(t)$.
\end{enumerate}
We now discuss each of these steps in detail.

\ssubsection{Pulse shape and noise spectrum}
\label{sec:mockdata}
When energy is deposited in the KID, the resonant frequency decreases rapidly and then relaxes back to the baseline $\bar f$. This can be read out by measuring the complex transmission coefficient of the resonator, $S_{21}$, while sweeping the input frequency across $\bar f$. During such a sweep, $S_{21}$ traces out a circle in the complex plane. When energy is deposited in the system, $S_{21}$ is suddenly displaced from its position on the circle. This displacement can be decomposed into the displacement towards the origin, called the dissipation quadrature, and the displacement tangent to the circle, called the frequency quadrature. While both quadratures contain information about the detector response, the signal in the frequency quadrature has a longer amplitude and lifetime \cite{Guo:2017ukv}, so we use only the frequency quadrature in this work.
A priori, the pulse shape $s(t)$ in the frequency quadrature is expected to have the form
\begin{equation}
    s(t) \approx A_0\left[e^{-t/\tau_\qp} - e^{-t/\tau_{\mathrm{rise}}}\right]
    ,
\end{equation}
where $\tau_{\mathrm{rise}} \sim \qty{10}{\micro\second}$ and $t_{\qp} \sim \qty{100}{\micro\second}$. However, for our purposes, it is critical to have a robust characterization of the actual detector response. We thus determine the pulse shape entirely empirically.

\begin{figure}
    \centering
    \includegraphics[width=0.49\columnwidth]{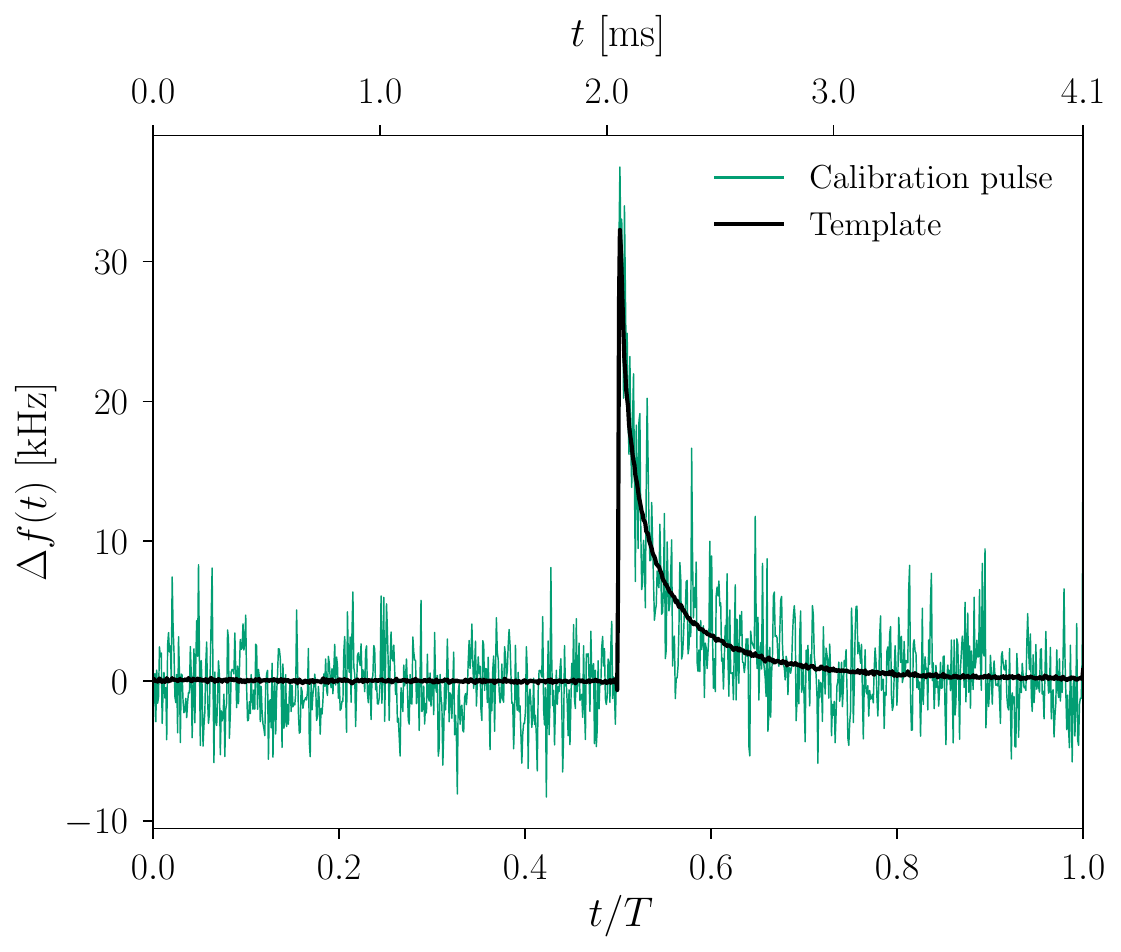}
    \hfill
    \includegraphics[width=0.49\columnwidth]{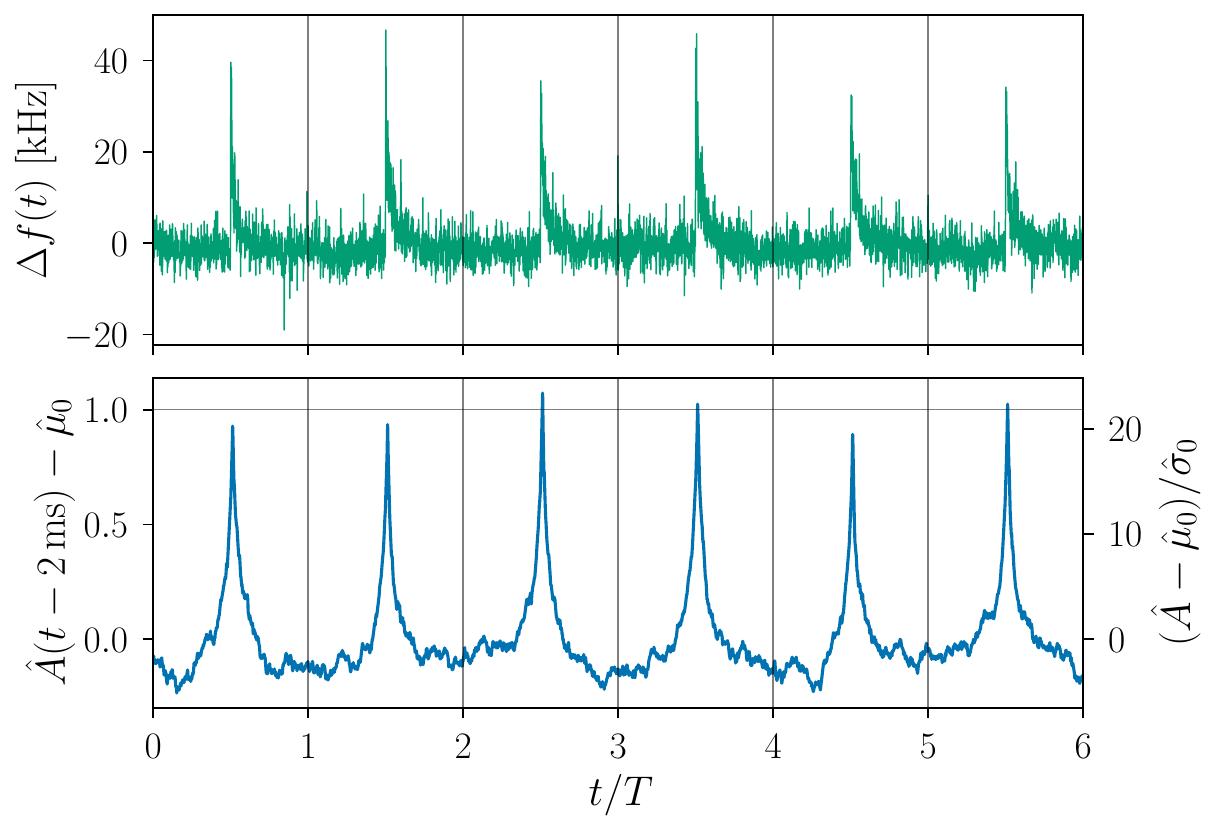}
    \caption{\textbf{Left:} Empirical template $s(t)$ obtained from calibration data, superimposed over data from a single calibration pulse. \textbf{Top right:} Raw timeseries for ten calibration pulses. \textbf{Bottom right:} Optimally-filtered timeseries, $\hat A$, shown with an offset of \qty{2}{\milli\second} so that peaks of $\hat A$ are aligned with peaks of $\Delta f$. The vertical axis on the right-hand side gives $\hat A$ in units of the width of the 0-photon peak, i.e., the width of the distribution of $\hat A$ on pure noise. The mean $\hat\mu_0$ of the 0-photon peak is subtracted. However, this has no qualitative effect, since $\hat\mu_0 \ll 0.01$.}
    \label{fig:template-and-filter}
\end{figure}

We analyze calibration data consisting of a timeseries of frequency quadrature measurements, $\Delta f(t)$, for \num{5000} laser pulses emitted at a rate of one per \qty{4.1}{\milli\second}. The laser wavelength is \qty{1550}{\nano\meter}, corresponding to photons with energy $\omega_c \equiv \qty{0.80}{\electronvolt}$. We divide the timeseries into \qty{4.1}{\milli\second} windows with a laser pulse emitted \qty{2}{\milli\second} after the beginning of each window. We thus assume that the first \qty{2}{\milli\second} of each window correspond to pure noise in addition to the baseline resonance frequency, i.e., $f(t) = \bar f + n(t)$. We average these data points across all \num{5000} windows to obtain the average resonance frequency,
\begin{equation}
    \bar{f} = \qty{1.50652}{\giga\hertz}
    .
\end{equation}
We likewise determine the pulse template $s(t)$ by averaging $\Delta f(t)$ over the \num{5000} windows. However, we do not take this average over every time series. The absorption efficiency is $\sim\!\! 1 / \mathrm{few}$, so not every window exhibits a pulse. We determine the pulse template by averaging only over those windows falling within the 1-photon peak, which we define to be those pulses whose maxima fall within one standard deviation in a Gaussian fit to the peak. This way, 1-photon pulses at the calibration wavelength correspond roughly to a pulse height $A=1$, and 2-photon pulses correspond roughly to a pulse height $A=2$. The resulting template is shown together with one window of calibration data in \cref{fig:template-and-filter}.

We use the pulse template $s(t)$ to identify events using a Wiener (optimal) filter. That is, given a sample with the same duration $T$ as the template, we estimate the pulse height $A$ with the estimator
\begin{equation}
    \label{eq:A-estimator}
    \hat A \equiv
    \frac{
        \sum_{k=-N/2}^{N/2-1}\tilde s_k^*\widetilde{\Delta f}_k\left|\tilde n_k\right|^{-2}
    }{
        \sum_{k=-N/2}^{N/2-1}
        \left|\tilde s_k\right|^2\left|\tilde n_k\right|^{-2}
    }
    ,
\end{equation}
where $v \equiv \Delta f$; $N$ is the number of samples recorded in time $T$; $f_k = k/T$; and $\tilde s_k$, $\widetilde{\Delta f}_k$, and $\tilde n_k$ are the discrete Fourier transforms of $s(t)$, $\Delta f(t)$, and $n(t)$, each evaluated at $f_k$. This estimator is optimal in the sense of minimizing $\chi^2$ (see e.g. \refcite{Golwala:2000zb}).

Now we return to the noise term $n(t)$. On general grounds, it is expected that $n(t)$ is approximately described by Gaussian white noise, i.e., $n(t)\sim\mathcal{N}(0, A_n)$. If this description is exact, then the noise spectrum does not influence the estimator $\hat A$ at all: that is, if $\tilde n$ is independent of frequency, the $|\tilde n_k|^2$ cancel between the numerator and denominator of \cref{eq:A-estimator}. However, the noise is not perfectly white. In particular, we find a $1/f$ excess below \qty{10}{\kilo\hertz} that is consistent with two-level noise, similar to that found in \refscite{Gao:2007,Gao:2012rb}. Thus, we need to obtain the full noise power spectrum empirically from calibration data. To that end, we concatenate the dark portions of all calibration windows ($t<\qty{2}{\milli\second}$), divide this timeseries into windows of length $T$, and compute $\tilde n_k$ on each window. We then average $|\tilde n_k|^2$ across these windows.

With the empirically-determined pulse shape and noise spectrum, we produce a filtered timeseries $\hat A(t)$ for the calibration data, where $\hat A(t)$ is defined to be the value of the estimator $\hat A$ on the window $(t, t+T)$. We verify that $\hat A \approx 1$ for pulses in the 1-photon peak, as shown in the right panel of \cref{fig:template-and-filter}. The calibration data also provides information about the detector response for larger deposits: in some fraction of events, two or more photons are simultaneously absorbed by the detector, leading to multiple peaks in the distribution of $\hat A$. We have checked from the calibration data that the same signal template holds for larger energy deposits, and that the mean signal amplitude $A$ scales linearly with the size of the deposit. Thus, an estimate of $A$ can be interpreted approximately as an estimate of $\omega / \omega_c$.

However, at fixed deposit size $\omega$, the value of $\hat A$ in calibration data is not perfectly fixed: rather, it is normally distributed with a nonnegligible width. Empirically, the parameters of this distribution are well approximated by making the following assumptions:
\begin{enumerate}
    \item The difference between the mean of the distribution, $\hat\mu_\omega$, and the mean of the 0-photon peak, $\hat\mu_0$, is linear in the deposit: that is, $\tilde\mu_\omega\equiv\hat\mu_\omega - \hat\mu_0\approx C_\mu\omega$.
    \item The width of the distribution $\hat\sigma_\omega$ receives two contributions: a width $\tilde\sigma_\omega = C_\sigma\tilde\mu_\omega$ that is linear in $\tilde\mu_\omega$, corresponding to deposit-induced noise, and the width $\hat\sigma_0$ of the 0-photon peak, corresponding to the baseline noise in the detector system. Combining these in quadrature yields $\hat\sigma_\omega \approx \sqrt{\hat\sigma_0^2 + C_\sigma^2\tilde\mu_\omega^2}$.
\end{enumerate}
Comparing the 0-, 1-, and 2-photon peaks in the calibration data, we find that the parameters of the peaks are well fit by $C_\mu = \qty{1.20}{\electronvolt^{-1}}$ and $C_\sigma = 0.16$. We use these parameters to perform a test of our event-counting pipeline by injecting a known number of sampled pulses into dark data and recovering them with our analysis procedure, as we detail in the next subsection.

\ssubsection{Event counting}
Once the timeseries $\hat A(t)$ is determined, it is necessary to define criteria by which events are identified and counted. There is no sharp intrinsic threshold in the KID detector apart from the superconducting gap: the threshold can in principle be reduced almost arbitrarily, at the cost of increasing the dark count rate. In practice, an ad-hoc threshold must be applied on $\hat A(t)$ to count energy deposition events, and this threshold determines the dark count rate. In calibration data, the estimator $\hat A$ has a well characterized distribution on pure noise, as shown by the 0-photon peak in \cref{fig:n-photon-peaks}. Thus, a fluctuation in $\hat A$ originating from noise can be assigned a statistical significance, as shown at right in \cref{fig:template-and-filter}. With a large sampling rate of \qty{250}{\kilo\hertz}, a high significance threshold is required to make the dark count rate manageable.

Given a threshold $\hat A_{\mathrm{min}}$, an event is counted when $\hat A$ exceeds $\hat A_{\mathrm{min}}$, and no further events are counted until $\hat A$ crosses below $\hat\sigma_0$, the width of the distribution of $\hat A$ in the absence of pulses. For $Z\equiv\hat A_{\mathrm{min}}/\hat\sigma_0 \gg 1$, if the distribution on $\hat A$ on noise is perfectly Gaussian, this leads to a dark count rate of $\Gamma_0 = (2\pi)^{-1/2}Z^{-1}e^{-\frac12 Z^2}\times\Gamma_{\mathrm s}$, where $\Gamma_{\mathrm s}$ is the sampling rate. That is,
\begin{equation}
    \Gamma_0 \approx \qty{4.1}{\year^{-1}}\times
        \left(\frac{\Gamma_{\mathrm s}}{\qty{250}{\kilo\hertz}}\right)
        \left(\frac{Z}{7}\right)^{-1}
        \exp\left[-\frac12\left(Z^2 - 7^2\right)\right]
        .
\end{equation}
Since $\hat A$ corresponds to the size of the deposit, $\hat A_{\mathrm{min}}$ corresponds directly to a threshold in deposited energy. The fiducial threshold $Z=7$ above corresponds to $\omega_{\mathrm{min}} \approx \qty{200}{\milli\electronvolt}$. However, as shown in \cref{fig:dark-count-rate}, our dataset contains excess dark counts at higher energies (larger $\hat A$) that cannot be accounted for by the noise distribution measured from calibration data. The estimate above is only valid if the sources that account for these events can be identified and eliminated.

\begin{figure}\centering
    \includegraphics[width=0.49\columnwidth]{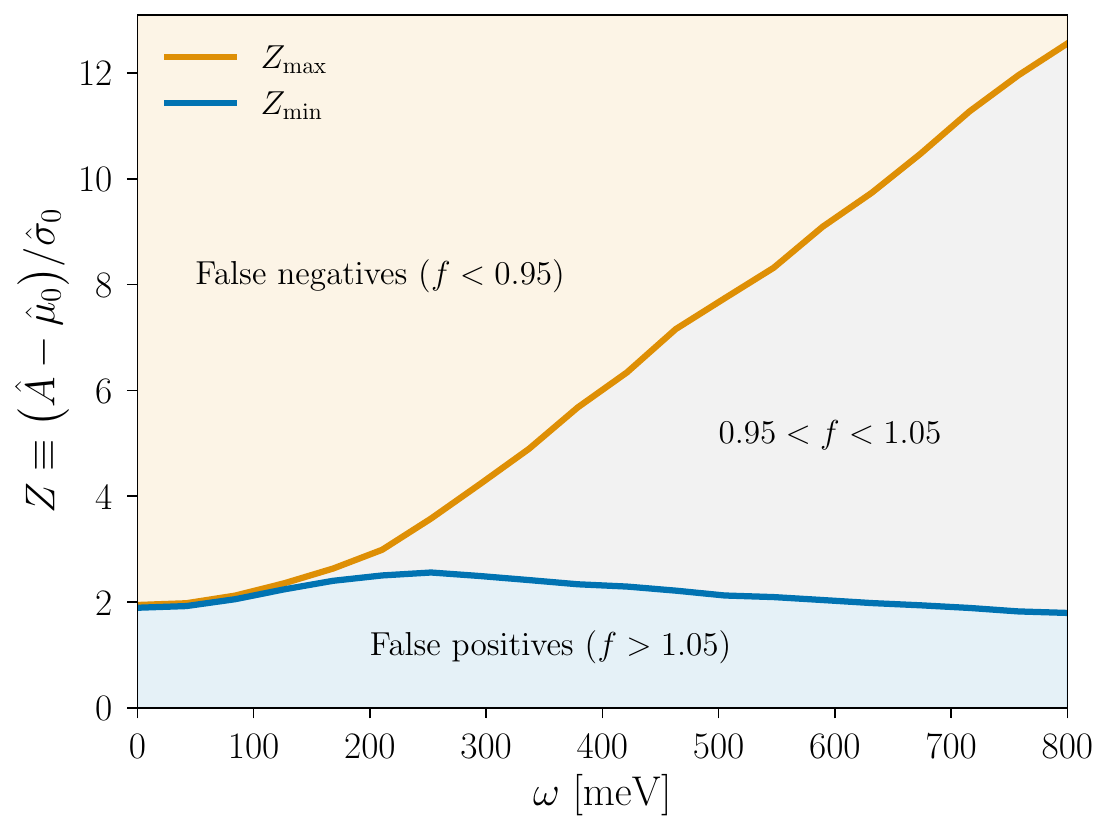}
    \caption{Fraction of simulated pulses reconstructed as a function of pulse energy and threshold. Thresholds are shown in units of the width of the 0-photon peak, i.e., the width of $\hat A$ on pure noise. In the orange region (high thresholds), fewer than 95\% of injected pulses are counted. In the blue region (low thresholds), the number of pulses counted is larger than the number injected by at least 5\%. In the gray region, the number of pulses counted is within 5\% of the number injected. The blue and orange curves show $Z_{\mathrm{min}}$ and $Z_{\mathrm{max}}$ as defined in the text.}
    \label{fig:thresholds}
\end{figure}

Another way to characterize the effective threshold is by demanding that our pipeline should be able to identify a number of injected events in mock data to within a certain level of precision. We carry out this test using the distributional parameters obtained in the previous subsection for the relationship between $\hat A$ and the size of a deposit. Assuming these distributions, we generate a sample of $N_{\mathrm{inj}}$ mock events at varying energies, count the number of events $N_{\mathrm{obs}}$ in the resulting timeseries, and compute the fraction $f \equiv N_{\mathrm{obs}} / N_{\mathrm{inj}}$. In the absence of noise, with perfect detection efficiency, we should have $f = 1$. Realistically, as the threshold is lowered, $f$ exceeds 1, since noise fluctuations are misinterpreted as pulses. As the threshold is raised, on the other hand, $f$ approaches zero. Now, as a function of energy, we identify the range of thresholds $(Z_{\mathrm{min}}, Z_{\mathrm{max}})$ for which $0.95 < f < 1.05$. This is shown as the gray shaded region in \cref{fig:thresholds}. The detector can only operate at this level of precision at thresholds between $Z_{\mathrm{min}}$ and $Z_{\mathrm{max}}$, so the lowest effective threshold in energy is the energy below which $Z_{\mathrm{min}}$ and $Z_{\mathrm{max}}$ nearly coincide. From the figure, it is apparent that this effective threshold indeed corresponds to $\omega_{\mathrm{min}} \approx \qty{200}{\milli\electronvolt}$.

\ssection{Dark matter sensitivity analysis}
The preceding section detailed the process of identifying and counting events in our KID device. In the present section, we give supplementary details of the dark matter models constrained in the main text, and highlight features of the statistical treatment entering the constraints and projections in \cref{fig:reach}.

We model the detector system as a dielectric, and compute DM interaction rates from the electronic response function, following \refscite{Hochberg:2021pkt,Knapen:2021run,Boyd:2022tcn}. We determine the energy- and momentum-dependent response function $\chi(\bb q, \omega)$ from the Drude model, in which $\epsilon(\omega) = 1 - \omega_\plas^2/\omega^2$, where the plasma frequency $\omega_\plas$ is a material-dependent property. We use $\omega_\plas = \qty{7}{\electronvolt}$ for TiN. We set $\chi = (\bb q^2/e^2)\Im(-1/\epsilon)$ and neglect boundary considerations that may arise from the thin-layer geometry of the detector \cite{Hochberg:2021yud,DeRocco:2022rze}. Given this response function for the detector, we can now compute DM scattering and absorption rates with mild assumptions on the structure of the DM interactions. In particular, our scattering computations are valid for any scalar or vector interaction, and our absorption computations are valid for the case of a kinetic mixing between a dark vector and the Standard Model photon. Since the models and statistical treatment are different between the two cases, we discuss absorption and scattering separately in the following subsections.

\ssubsection{Dark matter absorption}
We begin with DM absorption because the statistical treatment is simpler, but we note that since absorption takes place in the regime where $\omega\approx m_\dm$ rather than $\omega \ll m_\dm$, absorption rates are not model-independent in the same sense as scattering rates. Therefore, we study absorption under the assumption of a fiducial model in which the DM is a vector $A_\mu'$ (i.e., a dark photon) interacting with the Standard Model photon $A_\mu$ via a kinetic mixing. That is, we assume an interaction Lagrangian of the form $\mathcal L_{\mathrm{int}} = -\frac12\kappa F_{\mu\nu}F^{\prime\mu\nu}$, where $F_{\mu\nu}^{(\prime)} \equiv \partial_\mu A^{(\prime)}_\nu - \partial_\nu A^{(\prime)}_\mu$. In this case, the absorption rate per unit volume takes the form \cite{Hochberg:2021yud}
\begin{equation}
    \label{eq:absorption-rate}
    \Gamma_{\mathrm{A}} =
        m_\dm\frac{\kappa^2e^2}{\bb q^2}\chi(\bb p_\dm, m_\dm)\,,
\end{equation} 
so our experiment constrains the coupling $\kappa$ at fixed $m_\dm$.

The statistical treatment of DM absorption is relatively straightforward, since an absorption signal is monochromatic: to good approximation, all events deposit the same energy, $\omega \approx m_\dm$. Thus, at each DM mass, only one bin of $\omega$ is relevant, and it suffices to count events in that bin alone. In turn, this means that each value of $m_\dm$ is subject to a background rate limited to that same bin. Then a constraint can be set by the usual Feldman--Cousins procedure \cite{Feldman:1997qc}, simply on the basis of the total count rate in the corresponding bin. The difference between our analysis and a typical threshold-based analysis (e.g. as in \refcite{Hochberg:2021yud}) is that the threshold is set equal to $m_\dm$ for each DM mass, which means that the dark count rate is different for each $m_\dm$, becoming very large for $m_\dm \lesssim \qty{200}{\milli\electronvolt}$. Accordingly, our absorption constraint does not show a sharp cutoff at low masses, but relaxes smoothly as the threshold decreases and the dark count rate increases. Note that further background rejection could be achieved using the fact that the signal is a narrow line. This is not incorporated in our analysis.

\ssubsection{Dark matter scattering}
For DM scattering, the event rate is model-independent for spin-independent interactions (i.e., scalar or vector mediators), and takes the form given in \cref{eq:scattering-rate} of the main text:
\begin{equation}
    \label{eq:scattering-rate}
    \Gamma(\bb v_\dm) = \int\frac{\du^3\bb q}{(2\pi)^3}|V(\bb q)|^2\left[
        2\chi(\bb q, \omega_{\bb q})
    \right]
    \,.
\end{equation}
However, unlike absorption, scattering produces a broad spectrum of deposits whose shape is dependent on the mediator mass. The DM--electron interaction potential $V(\bb q)$ can be written in the nonrelativistic limit as $V(\bb q) = g_eg_\dm/(\bb q^2 + m_\med^2)$, so there are two distinct regimes: given a typical momentum transfer $q_{\mathrm{ref}} = \alpha_{\mathrm{EM}}m_e$, the light mediator regime $m_\med \ll q_{\mathrm{ref}}$ produces a spectrum which increases at small deposits, while the heavy mediator regime $m_\med \gg q_{\mathrm{ref}}$ produces a spectrum peaking at the largest kinematically-allowed deposits.

In each case, the broad spectrum of deposits means that there is no direct mapping between the DM mass and the dark count rate. Instead, given a spectrum of deposits, we constrain the DM parameters using the profile likelihood ratio test. Here, we assume that all events counted are due to backgrounds and not due to DM scattering events. We then test for a preference between the background-only model and background plus signal. For a measured spectrum $\{\Gamma_i\}$ at energies $\omega_i$, the likelihood ratio test statistic $\tau$ is given by
\begin{equation}
    \tfrac12\tau \equiv
        -\sum_i\ell\left(0;\,\omega_i,\Gamma_i\right) +
        \max_{g_eg_\dm}\left[
            \sum_i\ell\left(g_eg_\dm;\,\omega_i,\Gamma_i\right)
        \right]
    ,
\end{equation}
where $\ell\left(g_eg_\dm;\,\omega_i,\Gamma_i\right)$ is the log likelihood of the rate $\Gamma_i$ at energy $\omega_i$ at a fixed value of $g_eg_\dm$, such that $\ell\left(0;\,\omega_i,\Gamma_i\right)$ corresponds to the background-only hypothesis. We draw our constraint and projection curves at reference cross section $\bar\sigma_e$ corresponding to the value of $g_eg_\dm$ which produces a test statistic at the 95\% confidence level. In principle, this is not the most conservative constraint, since it assumes that the background model is robust. In our constraints and projections, we set the lower bound of our lowest bin at \qty{213}{\milli\electronvolt} and avoid any reliance on background modeling at lower energies.

\begin{figure}\centering
    \includegraphics[width=0.5\textwidth]{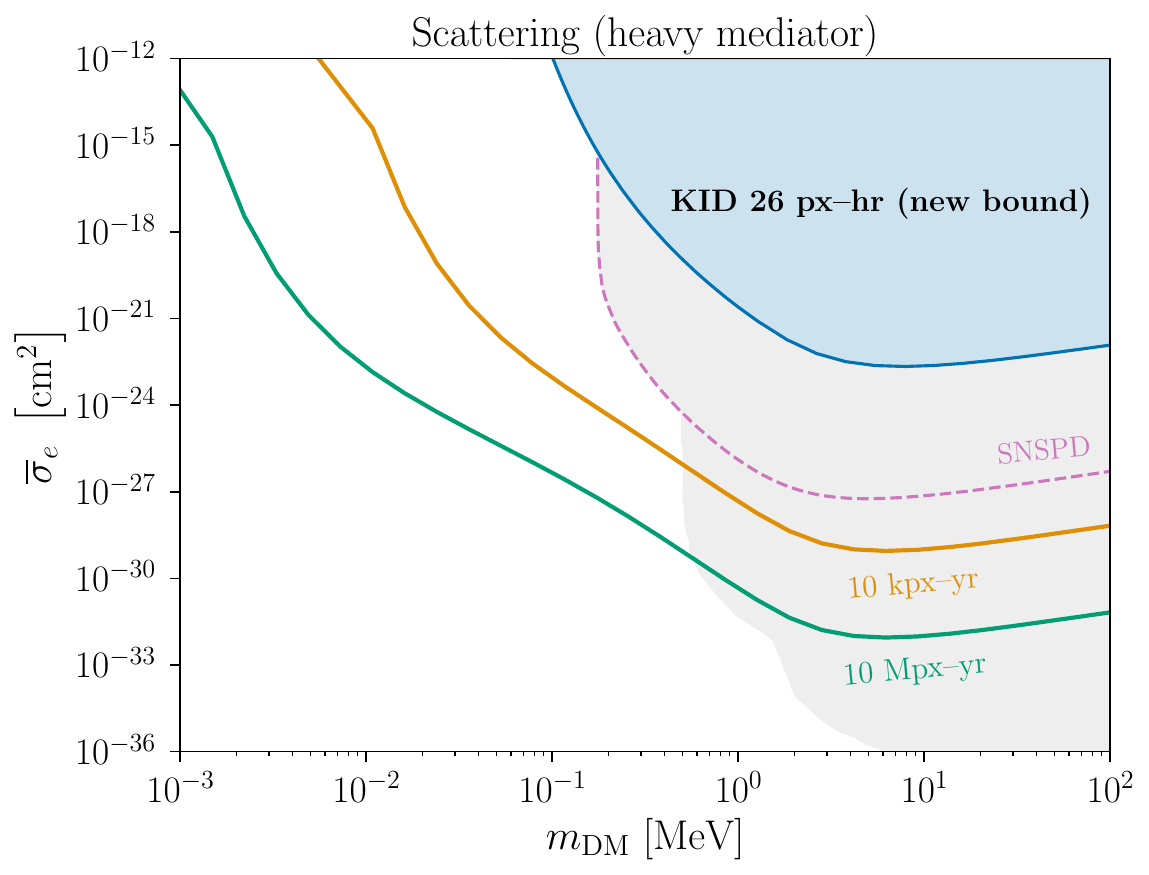}
    \caption{New bound and projections for DM scattering with electrons via a heavy mediator. All other features are shown as in the left panel of \cref{fig:reach} of the main text. The shaded gray region shows existing bounds from terrestrial experiments \cite{Barak:2020fql,Amaral:2020ryn,Aguilar-Arevalo:2019wdi,Essig:2017kqs,Agnes:2018oej,Aprile:2019xxb}, with recent SNSPD results outlined in dashed purple. The orange and green curves show projected constraints for year-long exposures of multiplexed KIDs with \num{e4} and \num{e7} pixels, respectively, each of the same size as our single-pixel prototype. In the projected curves, we assume that the high-energy dark counts in \cref{fig:dark-count-rate} are mitigated, such that background counts are produced only by the Gaussian white noise associated with the 0-photon peak. The width of this peak is reduced for the projected curves, corresponding to an increased energy resolving power and a decreased threshold. The resolving power for the orange curve is increased by a factor of 8, matching demonstrated resolving power in comparable devices. The green curve assumes the maximum resolving power in a low-gap absorber material with $T_C = \qty{100}{\milli\kelvin}$ ($R=136.2$).}
    \label{fig:heavy-mediator}
\end{figure}

We show constraints and projections for DM scattering via a light mediator in \cref{fig:reach} of the main text. For completeness, we show constraints and projections for scattering via a heavy mediator in \cref{fig:heavy-mediator}.

\bibliography{references}  

\begin{thebibliography}{51}%
\makeatletter
\providecommand \@ifxundefined [1]{%
 \@ifx{#1\undefined}
}%
\providecommand \@ifnum [1]{%
 \ifnum #1\expandafter \@firstoftwo
 \else \expandafter \@secondoftwo
 \fi
}%
\providecommand \@ifx [1]{%
 \ifx #1\expandafter \@firstoftwo
 \else \expandafter \@secondoftwo
 \fi
}%
\providecommand \natexlab [1]{#1}%
\providecommand \enquote  [1]{``#1''}%
\providecommand \bibnamefont  [1]{#1}%
\providecommand \bibfnamefont [1]{#1}%
\providecommand \citenamefont [1]{#1}%
\providecommand \href@noop [0]{\@secondoftwo}%
\providecommand \href [0]{\begingroup \@sanitize@url \@href}%
\providecommand \@href[1]{\@@startlink{#1}\@@href}%
\providecommand \@@href[1]{\endgroup#1\@@endlink}%
\providecommand \@sanitize@url [0]{\catcode `\\12\catcode `\$12\catcode
  `\&12\catcode `\#12\catcode `\^12\catcode `\_12\catcode `\%12\relax}%
\providecommand \@@startlink[1]{}%
\providecommand \@@endlink[0]{}%
\providecommand \url  [0]{\begingroup\@sanitize@url \@url }%
\providecommand \@url [1]{\endgroup\@href {#1}{\urlprefix }}%
\providecommand \urlprefix  [0]{URL }%
\providecommand \Eprint [0]{\href }%
\providecommand \doibase [0]{https://doi.org/}%
\providecommand \selectlanguage [0]{\@gobble}%
\providecommand \bibinfo  [0]{\@secondoftwo}%
\providecommand \bibfield  [0]{\@secondoftwo}%
\providecommand \translation [1]{[#1]}%
\providecommand \BibitemOpen [0]{}%
\providecommand \bibitemStop [0]{}%
\providecommand \bibitemNoStop [0]{.\EOS\space}%
\providecommand \EOS [0]{\spacefactor3000\relax}%
\providecommand \BibitemShut  [1]{\csname bibitem#1\endcsname}%
\let\auto@bib@innerbib\@empty
\bibitem [{\citenamefont {Essig}\ \emph {et~al.}(2016)\citenamefont {Essig},
  \citenamefont {Fernandez-Serra}, \citenamefont {Mardon}, \citenamefont
  {Soto}, \citenamefont {Volansky},\ and\ \citenamefont {Yu}}]{Essig:2015cda}%
  \BibitemOpen
  \bibfield  {author} {\bibinfo {author} {\bibfnamefont {R.}~\bibnamefont
  {Essig}}, \bibinfo {author} {\bibfnamefont {M.}~\bibnamefont
  {Fernandez-Serra}}, \bibinfo {author} {\bibfnamefont {J.}~\bibnamefont
  {Mardon}}, \bibinfo {author} {\bibfnamefont {A.}~\bibnamefont {Soto}},
  \bibinfo {author} {\bibfnamefont {T.}~\bibnamefont {Volansky}},\ and\
  \bibinfo {author} {\bibfnamefont {T.-T.}\ \bibnamefont {Yu}},\ }\bibfield
  {title} {\bibinfo {title} {{Direct Detection of sub-GeV Dark Matter with
  Semiconductor Targets}},\ }\href {https://doi.org/10.1007/JHEP05(2016)046}
  {\bibfield  {journal} {\bibinfo  {journal} {JHEP}\ }\textbf {\bibinfo
  {volume} {05}},\ \bibinfo {pages} {046}},\ \Eprint
  {https://arxiv.org/abs/1509.01598} {arXiv:1509.01598 [hep-ph]} \BibitemShut
  {NoStop}%
\bibitem [{\citenamefont {Graham}\ \emph {et~al.}(2012)\citenamefont {Graham},
  \citenamefont {Kaplan}, \citenamefont {Rajendran},\ and\ \citenamefont
  {Walters}}]{Graham:2012su}%
  \BibitemOpen
  \bibfield  {author} {\bibinfo {author} {\bibfnamefont {P.~W.}\ \bibnamefont
  {Graham}}, \bibinfo {author} {\bibfnamefont {D.~E.}\ \bibnamefont {Kaplan}},
  \bibinfo {author} {\bibfnamefont {S.}~\bibnamefont {Rajendran}},\ and\
  \bibinfo {author} {\bibfnamefont {M.~T.}\ \bibnamefont {Walters}},\
  }\bibfield  {title} {\bibinfo {title} {{Semiconductor Probes of Light Dark
  Matter}},\ }\href {https://doi.org/10.1016/j.dark.2012.09.001} {\bibfield
  {journal} {\bibinfo  {journal} {Phys. Dark Univ.}\ }\textbf {\bibinfo
  {volume} {1}},\ \bibinfo {pages} {32} (\bibinfo {year} {2012})},\ \Eprint
  {https://arxiv.org/abs/1203.2531} {arXiv:1203.2531 [hep-ph]} \BibitemShut
  {NoStop}%
\bibitem [{\citenamefont {Hochberg}\ \emph
  {et~al.}(2017{\natexlab{a}})\citenamefont {Hochberg}, \citenamefont {Lin},\
  and\ \citenamefont {Zurek}}]{Hochberg:2016sqx}%
  \BibitemOpen
  \bibfield  {author} {\bibinfo {author} {\bibfnamefont {Y.}~\bibnamefont
  {Hochberg}}, \bibinfo {author} {\bibfnamefont {T.}~\bibnamefont {Lin}},\ and\
  \bibinfo {author} {\bibfnamefont {K.~M.}\ \bibnamefont {Zurek}},\ }\bibfield
  {title} {\bibinfo {title} {{Absorption of light dark matter in
  semiconductors}},\ }\href {https://doi.org/10.1103/PhysRevD.95.023013}
  {\bibfield  {journal} {\bibinfo  {journal} {Phys. Rev.}\ }\textbf {\bibinfo
  {volume} {D95}},\ \bibinfo {pages} {023013} (\bibinfo {year}
  {2017}{\natexlab{a}})},\ \Eprint {https://arxiv.org/abs/1608.01994}
  {arXiv:1608.01994 [hep-ph]} \BibitemShut {NoStop}%
\bibitem [{\citenamefont {Griffin}\ \emph {et~al.}(2021)\citenamefont
  {Griffin}, \citenamefont {Hochberg}, \citenamefont {Inzani}, \citenamefont
  {Kurinsky}, \citenamefont {Lin},\ and\ \citenamefont {Yu}}]{Griffin:2020lgd}%
  \BibitemOpen
  \bibfield  {author} {\bibinfo {author} {\bibfnamefont {S.~M.}\ \bibnamefont
  {Griffin}}, \bibinfo {author} {\bibfnamefont {Y.}~\bibnamefont {Hochberg}},
  \bibinfo {author} {\bibfnamefont {K.}~\bibnamefont {Inzani}}, \bibinfo
  {author} {\bibfnamefont {N.}~\bibnamefont {Kurinsky}}, \bibinfo {author}
  {\bibfnamefont {T.}~\bibnamefont {Lin}},\ and\ \bibinfo {author}
  {\bibfnamefont {T.~C.}\ \bibnamefont {Yu}},\ }\bibfield  {title} {\bibinfo
  {title} {{Silicon carbide detectors for sub-GeV dark matter}},\ }\href
  {https://doi.org/10.1103/PhysRevD.103.075002} {\bibfield  {journal} {\bibinfo
   {journal} {Phys. Rev. D}\ }\textbf {\bibinfo {volume} {103}},\ \bibinfo
  {pages} {075002} (\bibinfo {year} {2021})},\ \Eprint
  {https://arxiv.org/abs/2008.08560} {arXiv:2008.08560 [hep-ph]} \BibitemShut
  {NoStop}%
\bibitem [{\citenamefont {Hochberg}\ \emph {et~al.}(2016)\citenamefont
  {Hochberg}, \citenamefont {Zhao},\ and\ \citenamefont
  {Zurek}}]{Hochberg:2015pha}%
  \BibitemOpen
  \bibfield  {author} {\bibinfo {author} {\bibfnamefont {Y.}~\bibnamefont
  {Hochberg}}, \bibinfo {author} {\bibfnamefont {Y.}~\bibnamefont {Zhao}},\
  and\ \bibinfo {author} {\bibfnamefont {K.~M.}\ \bibnamefont {Zurek}},\
  }\bibfield  {title} {\bibinfo {title} {{Superconducting Detectors for
  Superlight Dark Matter}},\ }\href
  {https://doi.org/10.1103/PhysRevLett.116.011301} {\bibfield  {journal}
  {\bibinfo  {journal} {Phys. Rev. Lett.}\ }\textbf {\bibinfo {volume} {116}},\
  \bibinfo {pages} {011301} (\bibinfo {year} {2016})},\ \Eprint
  {https://arxiv.org/abs/1504.07237} {arXiv:1504.07237 [hep-ph]} \BibitemShut
  {NoStop}%
\bibitem [{\citenamefont {Hochberg}\ \emph {et~al.}(2019)\citenamefont
  {Hochberg}, \citenamefont {Charaev}, \citenamefont {Nam}, \citenamefont
  {Verma}, \citenamefont {Colangelo},\ and\ \citenamefont
  {Berggren}}]{Hochberg:2019cyy}%
  \BibitemOpen
  \bibfield  {author} {\bibinfo {author} {\bibfnamefont {Y.}~\bibnamefont
  {Hochberg}}, \bibinfo {author} {\bibfnamefont {I.}~\bibnamefont {Charaev}},
  \bibinfo {author} {\bibfnamefont {S.-W.}\ \bibnamefont {Nam}}, \bibinfo
  {author} {\bibfnamefont {V.}~\bibnamefont {Verma}}, \bibinfo {author}
  {\bibfnamefont {M.}~\bibnamefont {Colangelo}},\ and\ \bibinfo {author}
  {\bibfnamefont {K.~K.}\ \bibnamefont {Berggren}},\ }\bibfield  {title}
  {\bibinfo {title} {{Detecting Sub-GeV Dark Matter with Superconducting
  Nanowires}},\ }\href {https://doi.org/10.1103/PhysRevLett.123.151802}
  {\bibfield  {journal} {\bibinfo  {journal} {Phys. Rev. Lett.}\ }\textbf
  {\bibinfo {volume} {123}},\ \bibinfo {pages} {151802} (\bibinfo {year}
  {2019})},\ \Eprint {https://arxiv.org/abs/1903.05101} {arXiv:1903.05101
  [hep-ph]} \BibitemShut {NoStop}%
\bibitem [{\citenamefont {Hochberg}\ \emph
  {et~al.}(2021{\natexlab{a}})\citenamefont {Hochberg}, \citenamefont
  {Lehmann}, \citenamefont {Charaev}, \citenamefont {Chiles}, \citenamefont
  {Colangelo}, \citenamefont {Nam},\ and\ \citenamefont
  {Berggren}}]{Hochberg:2021yud}%
  \BibitemOpen
  \bibfield  {author} {\bibinfo {author} {\bibfnamefont {Y.}~\bibnamefont
  {Hochberg}}, \bibinfo {author} {\bibfnamefont {B.~V.}\ \bibnamefont
  {Lehmann}}, \bibinfo {author} {\bibfnamefont {I.}~\bibnamefont {Charaev}},
  \bibinfo {author} {\bibfnamefont {J.}~\bibnamefont {Chiles}}, \bibinfo
  {author} {\bibfnamefont {M.}~\bibnamefont {Colangelo}}, \bibinfo {author}
  {\bibfnamefont {S.~W.}\ \bibnamefont {Nam}},\ and\ \bibinfo {author}
  {\bibfnamefont {K.~K.}\ \bibnamefont {Berggren}},\ }\bibfield  {title}
  {\bibinfo {title} {{New Constraints on Dark Matter from Superconducting
  Nanowires}},\ }\href@noop {} {\  (\bibinfo {year} {2021}{\natexlab{a}})},\
  \Eprint {https://arxiv.org/abs/2110.01586} {arXiv:2110.01586 [hep-ph]}
  \BibitemShut {NoStop}%
\bibitem [{\citenamefont {Hochberg}\ \emph
  {et~al.}(2017{\natexlab{b}})\citenamefont {Hochberg}, \citenamefont {Kahn},
  \citenamefont {Lisanti}, \citenamefont {Tully},\ and\ \citenamefont
  {Zurek}}]{Hochberg:2016ntt}%
  \BibitemOpen
  \bibfield  {author} {\bibinfo {author} {\bibfnamefont {Y.}~\bibnamefont
  {Hochberg}}, \bibinfo {author} {\bibfnamefont {Y.}~\bibnamefont {Kahn}},
  \bibinfo {author} {\bibfnamefont {M.}~\bibnamefont {Lisanti}}, \bibinfo
  {author} {\bibfnamefont {C.~G.}\ \bibnamefont {Tully}},\ and\ \bibinfo
  {author} {\bibfnamefont {K.~M.}\ \bibnamefont {Zurek}},\ }\bibfield  {title}
  {\bibinfo {title} {{Directional detection of dark matter with two-dimensional
  targets}},\ }\href {https://doi.org/10.1016/j.physletb.2017.06.051}
  {\bibfield  {journal} {\bibinfo  {journal} {Phys. Lett. B}\ }\textbf
  {\bibinfo {volume} {772}},\ \bibinfo {pages} {239} (\bibinfo {year}
  {2017}{\natexlab{b}})},\ \Eprint {https://arxiv.org/abs/1606.08849}
  {arXiv:1606.08849 [hep-ph]} \BibitemShut {NoStop}%
\bibitem [{\citenamefont {Cavoto}\ \emph {et~al.}(2018)\citenamefont {Cavoto},
  \citenamefont {Luchetta},\ and\ \citenamefont {Polosa}}]{Cavoto:2017otc}%
  \BibitemOpen
  \bibfield  {author} {\bibinfo {author} {\bibfnamefont {G.}~\bibnamefont
  {Cavoto}}, \bibinfo {author} {\bibfnamefont {F.}~\bibnamefont {Luchetta}},\
  and\ \bibinfo {author} {\bibfnamefont {A.~D.}\ \bibnamefont {Polosa}},\
  }\bibfield  {title} {\bibinfo {title} {{Sub-GeV Dark Matter Detection with
  Electron Recoils in Carbon Nanotubes}},\ }\href
  {https://doi.org/10.1016/j.physletb.2017.11.064} {\bibfield  {journal}
  {\bibinfo  {journal} {Phys. Lett. B}\ }\textbf {\bibinfo {volume} {776}},\
  \bibinfo {pages} {338} (\bibinfo {year} {2018})},\ \Eprint
  {https://arxiv.org/abs/1706.02487} {arXiv:1706.02487 [hep-ph]} \BibitemShut
  {NoStop}%
\bibitem [{\citenamefont {Schutz}\ and\ \citenamefont
  {Zurek}(2016)}]{Schutz:2016tid}%
  \BibitemOpen
  \bibfield  {author} {\bibinfo {author} {\bibfnamefont {K.}~\bibnamefont
  {Schutz}}\ and\ \bibinfo {author} {\bibfnamefont {K.~M.}\ \bibnamefont
  {Zurek}},\ }\bibfield  {title} {\bibinfo {title} {{Detectability of Light
  Dark Matter with Superfluid Helium}},\ }\href
  {https://doi.org/10.1103/PhysRevLett.117.121302} {\bibfield  {journal}
  {\bibinfo  {journal} {Phys. Rev. Lett.}\ }\textbf {\bibinfo {volume} {117}},\
  \bibinfo {pages} {121302} (\bibinfo {year} {2016})},\ \Eprint
  {https://arxiv.org/abs/1604.08206} {arXiv:1604.08206 [hep-ph]} \BibitemShut
  {NoStop}%
\bibitem [{\citenamefont {Knapen}\ \emph {et~al.}(2017)\citenamefont {Knapen},
  \citenamefont {Lin},\ and\ \citenamefont {Zurek}}]{Knapen:2016cue}%
  \BibitemOpen
  \bibfield  {author} {\bibinfo {author} {\bibfnamefont {S.}~\bibnamefont
  {Knapen}}, \bibinfo {author} {\bibfnamefont {T.}~\bibnamefont {Lin}},\ and\
  \bibinfo {author} {\bibfnamefont {K.~M.}\ \bibnamefont {Zurek}},\ }\bibfield
  {title} {\bibinfo {title} {{Light Dark Matter in Superfluid Helium: Detection
  with Multi-excitation Production}},\ }\href
  {https://doi.org/10.1103/PhysRevD.95.056019} {\bibfield  {journal} {\bibinfo
  {journal} {Phys. Rev.}\ }\textbf {\bibinfo {volume} {D95}},\ \bibinfo {pages}
  {056019} (\bibinfo {year} {2017})},\ \Eprint
  {https://arxiv.org/abs/1611.06228} {arXiv:1611.06228 [hep-ph]} \BibitemShut
  {NoStop}%
\bibitem [{\citenamefont {Hochberg}\ \emph {et~al.}(2018)\citenamefont
  {Hochberg}, \citenamefont {Kahn}, \citenamefont {Lisanti}, \citenamefont
  {Zurek}, \citenamefont {Grushin}, \citenamefont {Ilan}, \citenamefont
  {Griffin}, \citenamefont {Liu}, \citenamefont {Weber},\ and\ \citenamefont
  {Neaton}}]{Hochberg:2017wce}%
  \BibitemOpen
  \bibfield  {author} {\bibinfo {author} {\bibfnamefont {Y.}~\bibnamefont
  {Hochberg}}, \bibinfo {author} {\bibfnamefont {Y.}~\bibnamefont {Kahn}},
  \bibinfo {author} {\bibfnamefont {M.}~\bibnamefont {Lisanti}}, \bibinfo
  {author} {\bibfnamefont {K.~M.}\ \bibnamefont {Zurek}}, \bibinfo {author}
  {\bibfnamefont {A.~G.}\ \bibnamefont {Grushin}}, \bibinfo {author}
  {\bibfnamefont {R.}~\bibnamefont {Ilan}}, \bibinfo {author} {\bibfnamefont
  {S.~M.}\ \bibnamefont {Griffin}}, \bibinfo {author} {\bibfnamefont {Z.-F.}\
  \bibnamefont {Liu}}, \bibinfo {author} {\bibfnamefont {S.~F.}\ \bibnamefont
  {Weber}},\ and\ \bibinfo {author} {\bibfnamefont {J.~B.}\ \bibnamefont
  {Neaton}},\ }\bibfield  {title} {\bibinfo {title} {{Detection of sub-MeV Dark
  Matter with Three-Dimensional Dirac Materials}},\ }\href
  {https://doi.org/10.1103/PhysRevD.97.015004} {\bibfield  {journal} {\bibinfo
  {journal} {Phys. Rev. D}\ }\textbf {\bibinfo {volume} {97}},\ \bibinfo
  {pages} {015004} (\bibinfo {year} {2018})},\ \Eprint
  {https://arxiv.org/abs/1708.08929} {arXiv:1708.08929 [hep-ph]} \BibitemShut
  {NoStop}%
\bibitem [{\citenamefont {Oripov}\ \emph {et~al.}(2023)\citenamefont {Oripov},
  \citenamefont {Rampini}, \citenamefont {Allmaras}, \citenamefont {Shaw},
  \citenamefont {Nam}, \citenamefont {Korzh},\ and\ \citenamefont
  {McCaughan}}]{Oripov:2023yod}%
  \BibitemOpen
  \bibfield  {author} {\bibinfo {author} {\bibfnamefont {B.~G.}\ \bibnamefont
  {Oripov}}, \bibinfo {author} {\bibfnamefont {D.~S.}\ \bibnamefont {Rampini}},
  \bibinfo {author} {\bibfnamefont {J.}~\bibnamefont {Allmaras}}, \bibinfo
  {author} {\bibfnamefont {M.~D.}\ \bibnamefont {Shaw}}, \bibinfo {author}
  {\bibfnamefont {S.~W.}\ \bibnamefont {Nam}}, \bibinfo {author} {\bibfnamefont
  {B.}~\bibnamefont {Korzh}},\ and\ \bibinfo {author} {\bibfnamefont {A.~N.}\
  \bibnamefont {McCaughan}},\ }\bibfield  {title} {\bibinfo {title} {{A
  superconducting nanowire single-photon camera with 400,000 pixels}},\ }\href
  {https://doi.org/10.1038/s41586-023-06550-2} {\bibfield  {journal} {\bibinfo
  {journal} {Nature}\ }\textbf {\bibinfo {volume} {622}},\ \bibinfo {pages}
  {730} (\bibinfo {year} {2023})},\ \Eprint {https://arxiv.org/abs/2306.09473}
  {arXiv:2306.09473 [quant-ph]} \BibitemShut {NoStop}%
\bibitem [{\citenamefont {Mazin}\ \emph {et~al.}(2002)\citenamefont {Mazin},
  \citenamefont {Day}, \citenamefont {LeDuc}, \citenamefont {Vayonakis},\ and\
  \citenamefont {Zmuidzinas}}]{mazin2002superconducting}%
  \BibitemOpen
  \bibfield  {author} {\bibinfo {author} {\bibfnamefont {B.~A.}\ \bibnamefont
  {Mazin}}, \bibinfo {author} {\bibfnamefont {P.~K.}\ \bibnamefont {Day}},
  \bibinfo {author} {\bibfnamefont {H.~G.}\ \bibnamefont {LeDuc}}, \bibinfo
  {author} {\bibfnamefont {A.}~\bibnamefont {Vayonakis}},\ and\ \bibinfo
  {author} {\bibfnamefont {J.}~\bibnamefont {Zmuidzinas}},\ }\bibfield  {title}
  {\bibinfo {title} {Superconducting kinetic inductance photon detectors},\
  }in\ \href@noop {} {\emph {\bibinfo {booktitle} {Highly Innovative Space
  Telescope Concepts}}},\ Vol.\ \bibinfo {volume} {4849}\ (\bibinfo
  {organization} {SPIE},\ \bibinfo {year} {2002})\ pp.\ \bibinfo {pages}
  {283--293}\BibitemShut {NoStop}%
\bibitem [{\citenamefont {Day}\ \emph {et~al.}(2003)\citenamefont {Day},
  \citenamefont {LeDuc}, \citenamefont {Mazin}, \citenamefont {Vayonakis},\
  and\ \citenamefont {Zmuidzinas}}]{Day:2003}%
  \BibitemOpen
  \bibfield  {author} {\bibinfo {author} {\bibfnamefont {P.~K.}\ \bibnamefont
  {Day}}, \bibinfo {author} {\bibfnamefont {H.~G.}\ \bibnamefont {LeDuc}},
  \bibinfo {author} {\bibfnamefont {B.~A.}\ \bibnamefont {Mazin}}, \bibinfo
  {author} {\bibfnamefont {A.}~\bibnamefont {Vayonakis}},\ and\ \bibinfo
  {author} {\bibfnamefont {J.}~\bibnamefont {Zmuidzinas}},\ }\bibfield  {title}
  {\bibinfo {title} {A broadband superconducting detector suitable for use in
  large arrays},\ }\href {https://doi.org/10.1038/nature02037} {\bibfield
  {journal} {\bibinfo  {journal} {Nature}\ }\textbf {\bibinfo {volume} {425}},\
  \bibinfo {pages} {817} (\bibinfo {year} {2003})}\BibitemShut {NoStop}%
\bibitem [{\citenamefont {Zmuidzinas}(2012)}]{Zmuidzinas:2012}%
  \BibitemOpen
  \bibfield  {author} {\bibinfo {author} {\bibfnamefont {J.}~\bibnamefont
  {Zmuidzinas}},\ }\bibfield  {title} {\bibinfo {title} {Superconducting
  microresonators: Physics and applications},\ }\href
  {https://doi.org/10.1146/annurev-conmatphys-020911-125022} {\bibfield
  {journal} {\bibinfo  {journal} {Annual Review of Condensed Matter Physics}\
  }\textbf {\bibinfo {volume} {3}},\ \bibinfo {pages} {169} (\bibinfo {year}
  {2012})}\BibitemShut {NoStop}%
\bibitem [{\citenamefont {Gao}\ \emph {et~al.}(2012)\citenamefont {Gao} \emph
  {et~al.}}]{Gao:2012rb}%
  \BibitemOpen
  \bibfield  {author} {\bibinfo {author} {\bibfnamefont {J.}~\bibnamefont
  {Gao}} \emph {et~al.},\ }\bibfield  {title} {\bibinfo {title} {{A
  titanium-nitride near-infrared kinetic inductance photon-counting detector
  and its anomalous electrodynamics}},\ }\href
  {https://doi.org/10.1063/1.4756916} {\bibfield  {journal} {\bibinfo
  {journal} {Appl. Phys. Lett.}\ }\textbf {\bibinfo {volume} {101}},\ \bibinfo
  {pages} {142602} (\bibinfo {year} {2012})},\ \Eprint
  {https://arxiv.org/abs/1208.0871} {arXiv:1208.0871 [cond-mat.supr-con]}
  \BibitemShut {NoStop}%
\bibitem [{\citenamefont {Baselmans}(2012)}]{Baselmans:2012}%
  \BibitemOpen
  \bibfield  {author} {\bibinfo {author} {\bibfnamefont {J.}~\bibnamefont
  {Baselmans}},\ }\bibfield  {title} {\bibinfo {title} {Kinetic inductance
  detectors},\ }\href@noop {} {\bibfield  {journal} {\bibinfo  {journal}
  {Journal of Low Temperature Physics}\ }\textbf {\bibinfo {volume} {167}},\
  \bibinfo {pages} {292} (\bibinfo {year} {2012})}\BibitemShut {NoStop}%
\bibitem [{\citenamefont {Guo}\ \emph {et~al.}(2017)\citenamefont {Guo} \emph
  {et~al.}}]{Guo:2017ukv}%
  \BibitemOpen
  \bibfield  {author} {\bibinfo {author} {\bibfnamefont {W.}~\bibnamefont
  {Guo}} \emph {et~al.},\ }\bibfield  {title} {\bibinfo {title} {{Counting Near
  Infrared Photons with Microwave Kinetic Inductance Detectors}},\ }\href
  {https://doi.org/10.1063/1.4984134} {\bibfield  {journal} {\bibinfo
  {journal} {Appl. Phys. Lett.}\ }\textbf {\bibinfo {volume} {110}},\ \bibinfo
  {pages} {212601} (\bibinfo {year} {2017})},\ \Eprint
  {https://arxiv.org/abs/1702.07993} {arXiv:1702.07993 [physics.ins-det]}
  \BibitemShut {NoStop}%
\bibitem [{\citenamefont {Mazin}\ \emph {et~al.}(2019)\citenamefont {Mazin}
  \emph {et~al.}}]{Mazin:2019xkb}%
  \BibitemOpen
  \bibfield  {author} {\bibinfo {author} {\bibfnamefont {B.~A.}\ \bibnamefont
  {Mazin}} \emph {et~al.},\ }\bibfield  {title} {\bibinfo {title} {{Optical and
  Near-IR Microwave Kinetic Inductance Detectors (MKIDs) in the 2020s}},\
  }\href@noop {} {\  (\bibinfo {year} {2019})},\ \Eprint
  {https://arxiv.org/abs/1908.02775} {arXiv:1908.02775 [astro-ph.IM]}
  \BibitemShut {NoStop}%
\bibitem [{\citenamefont {Golwala}\ \emph {et~al.}(2008)\citenamefont
  {Golwala}, \citenamefont {Gao}, \citenamefont {Moore}, \citenamefont {Mazin},
  \citenamefont {Eckart}, \citenamefont {Bumble}, \citenamefont {Day},
  \citenamefont {LeDuc},\ and\ \citenamefont {Zmuidzinas}}]{Golwala:2008}%
  \BibitemOpen
  \bibfield  {author} {\bibinfo {author} {\bibfnamefont {S.}~\bibnamefont
  {Golwala}}, \bibinfo {author} {\bibfnamefont {J.}~\bibnamefont {Gao}},
  \bibinfo {author} {\bibfnamefont {D.}~\bibnamefont {Moore}}, \bibinfo
  {author} {\bibfnamefont {B.}~\bibnamefont {Mazin}}, \bibinfo {author}
  {\bibfnamefont {M.}~\bibnamefont {Eckart}}, \bibinfo {author} {\bibfnamefont
  {B.}~\bibnamefont {Bumble}}, \bibinfo {author} {\bibfnamefont
  {P.}~\bibnamefont {Day}}, \bibinfo {author} {\bibfnamefont {H.~G.}\
  \bibnamefont {LeDuc}},\ and\ \bibinfo {author} {\bibfnamefont
  {J.}~\bibnamefont {Zmuidzinas}},\ }\bibfield  {title} {\bibinfo {title} {A
  wimp dark matter detector using mkids},\ }\href
  {https://doi.org/10.1007/s10909-007-9687-0} {\bibfield  {journal} {\bibinfo
  {journal} {Journal of Low Temperature Physics}\ }\textbf {\bibinfo {volume}
  {151}},\ \bibinfo {pages} {550} (\bibinfo {year} {2008})}\BibitemShut
  {NoStop}%
\bibitem [{\citenamefont {Cruciani}\ \emph {et~al.}(2022)\citenamefont
  {Cruciani} \emph {et~al.}}]{Cruciani:2022mbb}%
  \BibitemOpen
  \bibfield  {author} {\bibinfo {author} {\bibfnamefont {A.}~\bibnamefont
  {Cruciani}} \emph {et~al.},\ }\bibfield  {title} {\bibinfo {title} {{BULLKID:
  Monolithic array of particle absorbers sensed by kinetic inductance
  detectors}},\ }\href {https://doi.org/10.1063/5.0128723} {\bibfield
  {journal} {\bibinfo  {journal} {Appl. Phys. Lett.}\ }\textbf {\bibinfo
  {volume} {121}},\ \bibinfo {pages} {213504} (\bibinfo {year} {2022})},\
  \Eprint {https://arxiv.org/abs/2209.14806} {arXiv:2209.14806
  [physics.ins-det]} \BibitemShut {NoStop}%
\bibitem [{\citenamefont {Pagnanini}\ \emph {et~al.}(2015)\citenamefont
  {Pagnanini} \emph {et~al.}}]{Pagnanini:2015ulo}%
  \BibitemOpen
  \bibfield  {author} {\bibinfo {author} {\bibfnamefont {L.}~\bibnamefont
  {Pagnanini}} \emph {et~al.},\ }\bibfield  {title} {\bibinfo {title} {{CALDER:
  Cryogenic Light Detector for rare event searches}},\ }\href
  {https://doi.org/10.22323/1.244.0076} {\bibfield  {journal} {\bibinfo
  {journal} {PoS}\ }\textbf {\bibinfo {volume} {NEUTEL2015}},\ \bibinfo {pages}
  {076} (\bibinfo {year} {2015})},\ \Eprint {https://arxiv.org/abs/1512.08901}
  {arXiv:1512.08901 [physics.ins-det]} \BibitemShut {NoStop}%
\bibitem [{\citenamefont {{Mazin}}\ \emph {et~al.}(2013)\citenamefont
  {{Mazin}}, \citenamefont {{Meeker}}, \citenamefont {{Strader}}, \citenamefont
  {{Szypryt}}, \citenamefont {{Marsden}}, \citenamefont {{van Eyken}},
  \citenamefont {{Duggan}}, \citenamefont {{Walter}}, \citenamefont
  {{Ulbricht}}, \citenamefont {{Johnson}}, \citenamefont {{Bumble}},
  \citenamefont {{O'Brien}},\ and\ \citenamefont
  {{Stoughton}}}]{2013PASP..125.1348M}%
  \BibitemOpen
  \bibfield  {author} {\bibinfo {author} {\bibfnamefont {B.~A.}\ \bibnamefont
  {{Mazin}}}, \bibinfo {author} {\bibfnamefont {S.~R.}\ \bibnamefont
  {{Meeker}}}, \bibinfo {author} {\bibfnamefont {M.~J.}\ \bibnamefont
  {{Strader}}}, \bibinfo {author} {\bibfnamefont {P.}~\bibnamefont
  {{Szypryt}}}, \bibinfo {author} {\bibfnamefont {D.}~\bibnamefont
  {{Marsden}}}, \bibinfo {author} {\bibfnamefont {J.~C.}\ \bibnamefont {{van
  Eyken}}}, \bibinfo {author} {\bibfnamefont {G.~E.}\ \bibnamefont {{Duggan}}},
  \bibinfo {author} {\bibfnamefont {A.~B.}\ \bibnamefont {{Walter}}}, \bibinfo
  {author} {\bibfnamefont {G.}~\bibnamefont {{Ulbricht}}}, \bibinfo {author}
  {\bibfnamefont {M.}~\bibnamefont {{Johnson}}}, \bibinfo {author}
  {\bibfnamefont {B.}~\bibnamefont {{Bumble}}}, \bibinfo {author}
  {\bibfnamefont {K.}~\bibnamefont {{O'Brien}}},\ and\ \bibinfo {author}
  {\bibfnamefont {C.}~\bibnamefont {{Stoughton}}},\ }\bibfield  {title}
  {\bibinfo {title} {{ARCONS: A 2024 Pixel Optical through Near-IR Cryogenic
  Imaging Spectrophotometer}},\ }\href {https://doi.org/10.1086/674013}
  {\bibfield  {journal} {\bibinfo  {journal} {\pasp}\ }\textbf {\bibinfo
  {volume} {125}},\ \bibinfo {pages} {1348} (\bibinfo {year} {2013})},\ \Eprint
  {https://arxiv.org/abs/1306.4674} {arXiv:1306.4674 [astro-ph.IM]}
  \BibitemShut {NoStop}%
\bibitem [{\citenamefont {{Meeker}}\ \emph {et~al.}(2018)\citenamefont
  {{Meeker}}, \citenamefont {{Mazin}}, \citenamefont {{Walter}}, \citenamefont
  {{Strader}}, \citenamefont {{Fruitwala}}, \citenamefont {{Bockstiegel}},
  \citenamefont {{Szypryt}}, \citenamefont {{Ulbricht}}, \citenamefont
  {{Coiffard}}, \citenamefont {{Bumble}}, \citenamefont {{Cancelo}},
  \citenamefont {{Zmuda}}, \citenamefont {{Treptow}}, \citenamefont {{Wilcer}},
  \citenamefont {{Collura}}, \citenamefont {{Dodkins}}, \citenamefont
  {{Lipartito}}, \citenamefont {{Zobrist}}, \citenamefont {{Bottom}},
  \citenamefont {{Shelton}}, \citenamefont {{Mawet}}, \citenamefont {{van
  Eyken}}, \citenamefont {{Vasisht}},\ and\ \citenamefont
  {{Serabyn}}}]{2018PASP..130f5001M}%
  \BibitemOpen
  \bibfield  {author} {\bibinfo {author} {\bibfnamefont {S.~R.}\ \bibnamefont
  {{Meeker}}}, \bibinfo {author} {\bibfnamefont {B.~A.}\ \bibnamefont
  {{Mazin}}}, \bibinfo {author} {\bibfnamefont {A.~B.}\ \bibnamefont
  {{Walter}}}, \bibinfo {author} {\bibfnamefont {P.}~\bibnamefont {{Strader}}},
  \bibinfo {author} {\bibfnamefont {N.}~\bibnamefont {{Fruitwala}}}, \bibinfo
  {author} {\bibfnamefont {C.}~\bibnamefont {{Bockstiegel}}}, \bibinfo {author}
  {\bibfnamefont {P.}~\bibnamefont {{Szypryt}}}, \bibinfo {author}
  {\bibfnamefont {G.}~\bibnamefont {{Ulbricht}}}, \bibinfo {author}
  {\bibfnamefont {G.}~\bibnamefont {{Coiffard}}}, \bibinfo {author}
  {\bibfnamefont {B.}~\bibnamefont {{Bumble}}}, \bibinfo {author}
  {\bibfnamefont {G.}~\bibnamefont {{Cancelo}}}, \bibinfo {author}
  {\bibfnamefont {T.}~\bibnamefont {{Zmuda}}}, \bibinfo {author} {\bibfnamefont
  {K.}~\bibnamefont {{Treptow}}}, \bibinfo {author} {\bibfnamefont
  {N.}~\bibnamefont {{Wilcer}}}, \bibinfo {author} {\bibfnamefont
  {G.}~\bibnamefont {{Collura}}}, \bibinfo {author} {\bibfnamefont
  {R.}~\bibnamefont {{Dodkins}}}, \bibinfo {author} {\bibfnamefont
  {I.}~\bibnamefont {{Lipartito}}}, \bibinfo {author} {\bibfnamefont
  {N.}~\bibnamefont {{Zobrist}}}, \bibinfo {author} {\bibfnamefont
  {M.}~\bibnamefont {{Bottom}}}, \bibinfo {author} {\bibfnamefont {J.~C.}\
  \bibnamefont {{Shelton}}}, \bibinfo {author} {\bibfnamefont {D.}~\bibnamefont
  {{Mawet}}}, \bibinfo {author} {\bibfnamefont {J.~C.}\ \bibnamefont {{van
  Eyken}}}, \bibinfo {author} {\bibfnamefont {G.}~\bibnamefont {{Vasisht}}},\
  and\ \bibinfo {author} {\bibfnamefont {E.}~\bibnamefont {{Serabyn}}},\
  }\bibfield  {title} {\bibinfo {title} {{DARKNESS: A Microwave Kinetic
  Inductance Detector Integral Field Spectrograph for High-contrast
  Astronomy}},\ }\href {https://doi.org/10.1088/1538-3873/aab5e7} {\bibfield
  {journal} {\bibinfo  {journal} {\pasp}\ }\textbf {\bibinfo {volume} {130}},\
  \bibinfo {pages} {065001} (\bibinfo {year} {2018})},\ \Eprint
  {https://arxiv.org/abs/1803.10420} {arXiv:1803.10420 [astro-ph.IM]}
  \BibitemShut {NoStop}%
\bibitem [{\citenamefont {{Walter}}\ \emph {et~al.}(2020)\citenamefont
  {{Walter}}, \citenamefont {{Fruitwala}}, \citenamefont {{Steiger}},
  \citenamefont {{Bailey}}, \citenamefont {{Zobrist}}, \citenamefont
  {{Swimmer}}, \citenamefont {{Lipartito}}, \citenamefont {{Smith}},
  \citenamefont {{Meeker}}, \citenamefont {{Bockstiegel}}, \citenamefont
  {{Coiffard}}, \citenamefont {{Dodkins}}, \citenamefont {{Szypryt}},
  \citenamefont {{Davis}}, \citenamefont {{Daal}}, \citenamefont {{Bumble}},
  \citenamefont {{Collura}}, \citenamefont {{Guyon}}, \citenamefont {{Lozi}},
  \citenamefont {{Vievard}}, \citenamefont {{Jovanovic}}, \citenamefont
  {{Martinache}}, \citenamefont {{Currie}},\ and\ \citenamefont
  {{Mazin}}}]{2020PASP..132l5005W}%
  \BibitemOpen
  \bibfield  {author} {\bibinfo {author} {\bibfnamefont {A.~B.}\ \bibnamefont
  {{Walter}}}, \bibinfo {author} {\bibfnamefont {N.}~\bibnamefont
  {{Fruitwala}}}, \bibinfo {author} {\bibfnamefont {S.}~\bibnamefont
  {{Steiger}}}, \bibinfo {author} {\bibfnamefont {I.}~\bibnamefont {{Bailey}},
  \bibfnamefont {John~I.}}, \bibinfo {author} {\bibfnamefont {N.}~\bibnamefont
  {{Zobrist}}}, \bibinfo {author} {\bibfnamefont {N.}~\bibnamefont
  {{Swimmer}}}, \bibinfo {author} {\bibfnamefont {I.}~\bibnamefont
  {{Lipartito}}}, \bibinfo {author} {\bibfnamefont {J.~P.}\ \bibnamefont
  {{Smith}}}, \bibinfo {author} {\bibfnamefont {S.~R.}\ \bibnamefont
  {{Meeker}}}, \bibinfo {author} {\bibfnamefont {C.}~\bibnamefont
  {{Bockstiegel}}}, \bibinfo {author} {\bibfnamefont {G.}~\bibnamefont
  {{Coiffard}}}, \bibinfo {author} {\bibfnamefont {R.}~\bibnamefont
  {{Dodkins}}}, \bibinfo {author} {\bibfnamefont {P.}~\bibnamefont
  {{Szypryt}}}, \bibinfo {author} {\bibfnamefont {K.~K.}\ \bibnamefont
  {{Davis}}}, \bibinfo {author} {\bibfnamefont {M.}~\bibnamefont {{Daal}}},
  \bibinfo {author} {\bibfnamefont {B.}~\bibnamefont {{Bumble}}}, \bibinfo
  {author} {\bibfnamefont {G.}~\bibnamefont {{Collura}}}, \bibinfo {author}
  {\bibfnamefont {O.}~\bibnamefont {{Guyon}}}, \bibinfo {author} {\bibfnamefont
  {J.}~\bibnamefont {{Lozi}}}, \bibinfo {author} {\bibfnamefont
  {S.}~\bibnamefont {{Vievard}}}, \bibinfo {author} {\bibfnamefont
  {N.}~\bibnamefont {{Jovanovic}}}, \bibinfo {author} {\bibfnamefont
  {F.}~\bibnamefont {{Martinache}}}, \bibinfo {author} {\bibfnamefont
  {T.}~\bibnamefont {{Currie}}},\ and\ \bibinfo {author} {\bibfnamefont
  {B.~A.}\ \bibnamefont {{Mazin}}},\ }\bibfield  {title} {\bibinfo {title}
  {{The MKID Exoplanet Camera for Subaru SCExAO}},\ }\href
  {https://doi.org/10.1088/1538-3873/abc60f} {\bibfield  {journal} {\bibinfo
  {journal} {\pasp}\ }\textbf {\bibinfo {volume} {132}},\ \bibinfo {eid}
  {125005} (\bibinfo {year} {2020})},\ \Eprint
  {https://arxiv.org/abs/2010.12620} {arXiv:2010.12620 [astro-ph.IM]}
  \BibitemShut {NoStop}%
\bibitem [{\citenamefont {O'Brien}(2020)}]{OBrien:2020uu}%
  \BibitemOpen
  \bibfield  {author} {\bibinfo {author} {\bibfnamefont {K.}~\bibnamefont
  {O'Brien}},\ }\bibfield  {title} {\bibinfo {title} {Kidspec: An mkid-based
  medium-resolution, integral field spectrograph},\ }\href
  {https://doi.org/10.1007/s10909-020-02347-z} {\bibfield  {journal} {\bibinfo
  {journal} {Journal of Low Temperature Physics}\ }\textbf {\bibinfo {volume}
  {199}},\ \bibinfo {pages} {537} (\bibinfo {year} {2020})}\BibitemShut
  {NoStop}%
\bibitem [{\citenamefont {McHugh}\ \emph {et~al.}(2012)\citenamefont {McHugh},
  \citenamefont {Mazin}, \citenamefont {Serfass}, \citenamefont {Meeker},
  \citenamefont {O'Brien}, \citenamefont {Duan}, \citenamefont {Raffanti},\
  and\ \citenamefont {Werthimer}}]{McHugh:2012dy}%
  \BibitemOpen
  \bibfield  {author} {\bibinfo {author} {\bibfnamefont {S.}~\bibnamefont
  {McHugh}}, \bibinfo {author} {\bibfnamefont {B.~A.}\ \bibnamefont {Mazin}},
  \bibinfo {author} {\bibfnamefont {B.}~\bibnamefont {Serfass}}, \bibinfo
  {author} {\bibfnamefont {S.}~\bibnamefont {Meeker}}, \bibinfo {author}
  {\bibfnamefont {K.}~\bibnamefont {O'Brien}}, \bibinfo {author} {\bibfnamefont
  {R.}~\bibnamefont {Duan}}, \bibinfo {author} {\bibfnamefont {R.}~\bibnamefont
  {Raffanti}},\ and\ \bibinfo {author} {\bibfnamefont {D.}~\bibnamefont
  {Werthimer}},\ }\bibfield  {title} {\bibinfo {title} {{A readout for large
  arrays of Microwave Kinetic Inductance Detectors}},\ }\href
  {https://doi.org/10.1063/1.3700812} {\bibfield  {journal} {\bibinfo
  {journal} {Rev. Sci. Instrum.}\ }\textbf {\bibinfo {volume} {83}},\ \bibinfo
  {pages} {044702} (\bibinfo {year} {2012})},\ \Eprint
  {https://arxiv.org/abs/1203.5861} {arXiv:1203.5861 [astro-ph.IM]}
  \BibitemShut {NoStop}%
\bibitem [{\citenamefont {Fruitwala}\ \emph {et~al.}(2020)\citenamefont
  {Fruitwala} \emph {et~al.}}]{Fruitwala:2020ibn}%
  \BibitemOpen
  \bibfield  {author} {\bibinfo {author} {\bibfnamefont {N.}~\bibnamefont
  {Fruitwala}} \emph {et~al.},\ }\bibfield  {title} {\bibinfo {title} {{Second
  Generation Readout For Large Format Photon Counting Microwave Kinetic
  Inductance Detectors}},\ }\href {https://doi.org/10.1063/5.0029457}
  {\bibfield  {journal} {\bibinfo  {journal} {Rev. Sci. Instrum.}\ }\textbf
  {\bibinfo {volume} {91}},\ \bibinfo {pages} {124705} (\bibinfo {year}
  {2020})},\ \Eprint {https://arxiv.org/abs/2011.06685} {arXiv:2011.06685
  [astro-ph.IM]} \BibitemShut {NoStop}%
\bibitem [{\citenamefont {Sinclair}\ \emph {et~al.}(2022)\citenamefont
  {Sinclair} \emph {et~al.}}]{CCAT-prime:2022tho}%
  \BibitemOpen
  \bibfield  {author} {\bibinfo {author} {\bibfnamefont {A.~K.}\ \bibnamefont
  {Sinclair}} \emph {et~al.} (\bibinfo {collaboration} {CCAT-prime}),\
  }\bibfield  {title} {\bibinfo {title} {{CCAT-prime: RFSoC based readout for
  frequency multiplexed kinetic inductance detectors}},\ }\href
  {https://doi.org/10.1117/12.2629722} {\bibfield  {journal} {\bibinfo
  {journal} {Proc. SPIE Int. Soc. Opt. Eng.}\ }\textbf {\bibinfo {volume}
  {12190}},\ \bibinfo {pages} {444} (\bibinfo {year} {2022})},\ \Eprint
  {https://arxiv.org/abs/2208.07465} {arXiv:2208.07465 [astro-ph.IM]}
  \BibitemShut {NoStop}%
\bibitem [{\citenamefont {Gao}(2008)}]{GaoThesis:2008}%
  \BibitemOpen
  \bibfield  {author} {\bibinfo {author} {\bibfnamefont {J.}~\bibnamefont
  {Gao}},\ }\bibfield  {title} {\bibinfo {title} {The physics of
  superconducting microwave resonators}} (\bibinfo {year} {2008})\BibitemShut
  {NoStop}%
\bibitem [{\citenamefont {Barak}\ \emph {et~al.}(2020)\citenamefont {Barak}
  \emph {et~al.}}]{Barak:2020fql}%
  \BibitemOpen
  \bibfield  {author} {\bibinfo {author} {\bibfnamefont {L.}~\bibnamefont
  {Barak}} \emph {et~al.} (\bibinfo {collaboration} {SENSEI}),\ }\bibfield
  {title} {\bibinfo {title} {{SENSEI: Direct-Detection Results on sub-GeV Dark
  Matter from a New Skipper-CCD}},\ }\href
  {https://doi.org/10.1103/PhysRevLett.125.171802} {\bibfield  {journal}
  {\bibinfo  {journal} {Phys. Rev. Lett.}\ }\textbf {\bibinfo {volume} {125}},\
  \bibinfo {pages} {171802} (\bibinfo {year} {2020})},\ \Eprint
  {https://arxiv.org/abs/2004.11378} {arXiv:2004.11378 [astro-ph.CO]}
  \BibitemShut {NoStop}%
\bibitem [{\citenamefont {Amaral}\ \emph {et~al.}(2020)\citenamefont {Amaral}
  \emph {et~al.}}]{Amaral:2020ryn}%
  \BibitemOpen
  \bibfield  {author} {\bibinfo {author} {\bibfnamefont {D.~W.}\ \bibnamefont
  {Amaral}} \emph {et~al.} (\bibinfo {collaboration} {SuperCDMS}),\ }\bibfield
  {title} {\bibinfo {title} {{Constraints on low-mass, relic dark matter
  candidates from a surface-operated SuperCDMS single-charge sensitive
  detector}},\ }\href {https://doi.org/10.1103/PhysRevD.102.091101} {\bibfield
  {journal} {\bibinfo  {journal} {Phys. Rev. D}\ }\textbf {\bibinfo {volume}
  {102}},\ \bibinfo {pages} {091101} (\bibinfo {year} {2020})},\ \Eprint
  {https://arxiv.org/abs/2005.14067} {arXiv:2005.14067 [hep-ex]} \BibitemShut
  {NoStop}%
\bibitem [{\citenamefont {Aguilar-Arevalo}\ \emph {et~al.}(2019)\citenamefont
  {Aguilar-Arevalo} \emph {et~al.}}]{Aguilar-Arevalo:2019wdi}%
  \BibitemOpen
  \bibfield  {author} {\bibinfo {author} {\bibfnamefont {A.}~\bibnamefont
  {Aguilar-Arevalo}} \emph {et~al.} (\bibinfo {collaboration} {DAMIC}),\
  }\bibfield  {title} {\bibinfo {title} {{Constraints on Light Dark Matter
  Particles Interacting with Electrons from DAMIC at SNOLAB}},\ }\href
  {https://doi.org/10.1103/PhysRevLett.123.181802} {\bibfield  {journal}
  {\bibinfo  {journal} {Phys. Rev. Lett.}\ }\textbf {\bibinfo {volume} {123}},\
  \bibinfo {pages} {181802} (\bibinfo {year} {2019})},\ \Eprint
  {https://arxiv.org/abs/1907.12628} {arXiv:1907.12628 [astro-ph.CO]}
  \BibitemShut {NoStop}%
\bibitem [{\citenamefont {Essig}\ \emph {et~al.}(2017)\citenamefont {Essig},
  \citenamefont {Volansky},\ and\ \citenamefont {Yu}}]{Essig:2017kqs}%
  \BibitemOpen
  \bibfield  {author} {\bibinfo {author} {\bibfnamefont {R.}~\bibnamefont
  {Essig}}, \bibinfo {author} {\bibfnamefont {T.}~\bibnamefont {Volansky}},\
  and\ \bibinfo {author} {\bibfnamefont {T.-T.}\ \bibnamefont {Yu}},\
  }\bibfield  {title} {\bibinfo {title} {{New Constraints and Prospects for
  sub-GeV Dark Matter Scattering off Electrons in Xenon}},\ }\href
  {https://doi.org/10.1103/PhysRevD.96.043017} {\bibfield  {journal} {\bibinfo
  {journal} {Phys. Rev. D}\ }\textbf {\bibinfo {volume} {96}},\ \bibinfo
  {pages} {043017} (\bibinfo {year} {2017})},\ \Eprint
  {https://arxiv.org/abs/1703.00910} {arXiv:1703.00910 [hep-ph]} \BibitemShut
  {NoStop}%
\bibitem [{\citenamefont {Agnes}\ \emph {et~al.}(2018)\citenamefont {Agnes}
  \emph {et~al.}}]{Agnes:2018oej}%
  \BibitemOpen
  \bibfield  {author} {\bibinfo {author} {\bibfnamefont {P.}~\bibnamefont
  {Agnes}} \emph {et~al.} (\bibinfo {collaboration} {DarkSide}),\ }\bibfield
  {title} {\bibinfo {title} {{Constraints on Sub-GeV
  Dark-Matter\textendash{}Electron Scattering from the DarkSide-50
  Experiment}},\ }\href {https://doi.org/10.1103/PhysRevLett.121.111303}
  {\bibfield  {journal} {\bibinfo  {journal} {Phys. Rev. Lett.}\ }\textbf
  {\bibinfo {volume} {121}},\ \bibinfo {pages} {111303} (\bibinfo {year}
  {2018})},\ \Eprint {https://arxiv.org/abs/1802.06998} {arXiv:1802.06998
  [astro-ph.CO]} \BibitemShut {NoStop}%
\bibitem [{\citenamefont {Aprile}\ \emph {et~al.}(2019)\citenamefont {Aprile}
  \emph {et~al.}}]{Aprile:2019xxb}%
  \BibitemOpen
  \bibfield  {author} {\bibinfo {author} {\bibfnamefont {E.}~\bibnamefont
  {Aprile}} \emph {et~al.} (\bibinfo {collaboration} {XENON}),\ }\bibfield
  {title} {\bibinfo {title} {{Light Dark Matter Search with Ionization Signals
  in XENON1T}},\ }\href {https://doi.org/10.1103/PhysRevLett.123.251801}
  {\bibfield  {journal} {\bibinfo  {journal} {Phys. Rev. Lett.}\ }\textbf
  {\bibinfo {volume} {123}},\ \bibinfo {pages} {251801} (\bibinfo {year}
  {2019})},\ \Eprint {https://arxiv.org/abs/1907.11485} {arXiv:1907.11485
  [hep-ex]} \BibitemShut {NoStop}%
\bibitem [{\citenamefont {An}\ \emph {et~al.}(2015)\citenamefont {An},
  \citenamefont {Pospelov}, \citenamefont {Pradler},\ and\ \citenamefont
  {Ritz}}]{An:2014twa}%
  \BibitemOpen
  \bibfield  {author} {\bibinfo {author} {\bibfnamefont {H.}~\bibnamefont
  {An}}, \bibinfo {author} {\bibfnamefont {M.}~\bibnamefont {Pospelov}},
  \bibinfo {author} {\bibfnamefont {J.}~\bibnamefont {Pradler}},\ and\ \bibinfo
  {author} {\bibfnamefont {A.}~\bibnamefont {Ritz}},\ }\bibfield  {title}
  {\bibinfo {title} {{Direct Detection Constraints on Dark Photon Dark
  Matter}},\ }\href {https://doi.org/10.1016/j.physletb.2015.06.018} {\bibfield
   {journal} {\bibinfo  {journal} {Phys. Lett. B}\ }\textbf {\bibinfo {volume}
  {747}},\ \bibinfo {pages} {331} (\bibinfo {year} {2015})},\ \Eprint
  {https://arxiv.org/abs/1412.8378} {arXiv:1412.8378 [hep-ph]} \BibitemShut
  {NoStop}%
\bibitem [{\citenamefont {Agnese}\ \emph {et~al.}(2018)\citenamefont {Agnese}
  \emph {et~al.}}]{Agnese:2018col}%
  \BibitemOpen
  \bibfield  {author} {\bibinfo {author} {\bibfnamefont {R.}~\bibnamefont
  {Agnese}} \emph {et~al.} (\bibinfo {collaboration} {SuperCDMS}),\ }\bibfield
  {title} {\bibinfo {title} {{First Dark Matter Constraints from a SuperCDMS
  Single-Charge Sensitive Detector}},\ }\href
  {https://doi.org/10.1103/PhysRevLett.121.051301} {\bibfield  {journal}
  {\bibinfo  {journal} {Phys. Rev. Lett.}\ }\textbf {\bibinfo {volume} {121}},\
  \bibinfo {pages} {051301} (\bibinfo {year} {2018})},\ \bibinfo {note}
  {[Erratum: Phys.Rev.Lett. 122, 069901 (2019)]},\ \Eprint
  {https://arxiv.org/abs/1804.10697} {arXiv:1804.10697 [hep-ex]} \BibitemShut
  {NoStop}%
\bibitem [{\citenamefont {Arnaud}\ \emph {et~al.}(2020)\citenamefont {Arnaud}
  \emph {et~al.}}]{Arnaud:2020svb}%
  \BibitemOpen
  \bibfield  {author} {\bibinfo {author} {\bibfnamefont {Q.}~\bibnamefont
  {Arnaud}} \emph {et~al.} (\bibinfo {collaboration} {EDELWEISS}),\ }\bibfield
  {title} {\bibinfo {title} {{First germanium-based constraints on sub-MeV Dark
  Matter with the EDELWEISS experiment}},\ }\href
  {https://doi.org/10.1103/PhysRevLett.125.141301} {\bibfield  {journal}
  {\bibinfo  {journal} {Phys. Rev. Lett.}\ }\textbf {\bibinfo {volume} {125}},\
  \bibinfo {pages} {141301} (\bibinfo {year} {2020})},\ \Eprint
  {https://arxiv.org/abs/2003.01046} {arXiv:2003.01046 [astro-ph.GA]}
  \BibitemShut {NoStop}%
\bibitem [{\citenamefont {Andrianavalomahefa}\ \emph
  {et~al.}(2020)\citenamefont {Andrianavalomahefa} \emph
  {et~al.}}]{FUNKExperiment:2020ofv}%
  \BibitemOpen
  \bibfield  {author} {\bibinfo {author} {\bibfnamefont {A.}~\bibnamefont
  {Andrianavalomahefa}} \emph {et~al.} (\bibinfo {collaboration} {FUNK
  Experiment}),\ }\bibfield  {title} {\bibinfo {title} {{Limits from the Funk
  Experiment on the Mixing Strength of Hidden-Photon Dark Matter in the Visible
  and Near-Ultraviolet Wavelength Range}},\ }\href
  {https://doi.org/10.1103/PhysRevD.102.042001} {\bibfield  {journal} {\bibinfo
   {journal} {Phys. Rev. D}\ }\textbf {\bibinfo {volume} {102}},\ \bibinfo
  {pages} {042001} (\bibinfo {year} {2020})},\ \Eprint
  {https://arxiv.org/abs/2003.13144} {arXiv:2003.13144 [astro-ph.CO]}
  \BibitemShut {NoStop}%
\bibitem [{\citenamefont {An}\ \emph {et~al.}(2013)\citenamefont {An},
  \citenamefont {Pospelov},\ and\ \citenamefont {Pradler}}]{An:2013yua}%
  \BibitemOpen
  \bibfield  {author} {\bibinfo {author} {\bibfnamefont {H.}~\bibnamefont
  {An}}, \bibinfo {author} {\bibfnamefont {M.}~\bibnamefont {Pospelov}},\ and\
  \bibinfo {author} {\bibfnamefont {J.}~\bibnamefont {Pradler}},\ }\bibfield
  {title} {\bibinfo {title} {{Dark Matter Detectors as Dark Photon
  Helioscopes}},\ }\href {https://doi.org/10.1103/PhysRevLett.111.041302}
  {\bibfield  {journal} {\bibinfo  {journal} {Phys. Rev. Lett.}\ }\textbf
  {\bibinfo {volume} {111}},\ \bibinfo {pages} {041302} (\bibinfo {year}
  {2013})},\ \Eprint {https://arxiv.org/abs/1304.3461} {arXiv:1304.3461
  [hep-ph]} \BibitemShut {NoStop}%
\bibitem [{\citenamefont {An}\ \emph {et~al.}(2020)\citenamefont {An},
  \citenamefont {Pospelov}, \citenamefont {Pradler},\ and\ \citenamefont
  {Ritz}}]{An:2020bxd}%
  \BibitemOpen
  \bibfield  {author} {\bibinfo {author} {\bibfnamefont {H.}~\bibnamefont
  {An}}, \bibinfo {author} {\bibfnamefont {M.}~\bibnamefont {Pospelov}},
  \bibinfo {author} {\bibfnamefont {J.}~\bibnamefont {Pradler}},\ and\ \bibinfo
  {author} {\bibfnamefont {A.}~\bibnamefont {Ritz}},\ }\bibfield  {title}
  {\bibinfo {title} {{New limits on dark photons from solar emission and keV
  scale dark matter}},\ }\href {https://doi.org/10.1103/PhysRevD.102.115022}
  {\bibfield  {journal} {\bibinfo  {journal} {Phys. Rev. D}\ }\textbf {\bibinfo
  {volume} {102}},\ \bibinfo {pages} {115022} (\bibinfo {year} {2020})},\
  \Eprint {https://arxiv.org/abs/2006.13929} {arXiv:2006.13929 [hep-ph]}
  \BibitemShut {NoStop}%
\bibitem [{\citenamefont {Hochberg}\ \emph
  {et~al.}(2021{\natexlab{b}})\citenamefont {Hochberg}, \citenamefont {Kahn},
  \citenamefont {Kurinsky}, \citenamefont {Lehmann}, \citenamefont {Yu},\ and\
  \citenamefont {Berggren}}]{Hochberg:2021pkt}%
  \BibitemOpen
  \bibfield  {author} {\bibinfo {author} {\bibfnamefont {Y.}~\bibnamefont
  {Hochberg}}, \bibinfo {author} {\bibfnamefont {Y.}~\bibnamefont {Kahn}},
  \bibinfo {author} {\bibfnamefont {N.}~\bibnamefont {Kurinsky}}, \bibinfo
  {author} {\bibfnamefont {B.~V.}\ \bibnamefont {Lehmann}}, \bibinfo {author}
  {\bibfnamefont {T.~C.}\ \bibnamefont {Yu}},\ and\ \bibinfo {author}
  {\bibfnamefont {K.~K.}\ \bibnamefont {Berggren}},\ }\bibfield  {title}
  {\bibinfo {title} {{Determining Dark Matter-Electron Scattering Rates from
  the Dielectric Function}},\ }\href@noop {} {\  (\bibinfo {year}
  {2021}{\natexlab{b}})},\ \Eprint {https://arxiv.org/abs/2101.08263}
  {arXiv:2101.08263 [hep-ph]} \BibitemShut {NoStop}%
\bibitem [{\citenamefont {de~Visser}\ \emph {et~al.}(2021)\citenamefont
  {de~Visser}, \citenamefont {de~Rooij}, \citenamefont {Murugesan},
  \citenamefont {Thoen},\ and\ \citenamefont {Baselmans}}]{deVisser:2021kip}%
  \BibitemOpen
  \bibfield  {author} {\bibinfo {author} {\bibfnamefont {P.~J.}\ \bibnamefont
  {de~Visser}}, \bibinfo {author} {\bibfnamefont {S.~A.~H.}\ \bibnamefont
  {de~Rooij}}, \bibinfo {author} {\bibfnamefont {V.}~\bibnamefont {Murugesan}},
  \bibinfo {author} {\bibfnamefont {D.~J.}\ \bibnamefont {Thoen}},\ and\
  \bibinfo {author} {\bibfnamefont {J.~J.~A.}\ \bibnamefont {Baselmans}},\
  }\bibfield  {title} {\bibinfo {title} {{Phonon-trapping enhanced energy
  resolution in superconducting single photon detectors}},\ }\href
  {https://doi.org/10.1103/PhysRevApplied.16.034051} {\bibfield  {journal}
  {\bibinfo  {journal} {Phys. Rev. Applied}\ }\textbf {\bibinfo {volume}
  {16}},\ \bibinfo {pages} {034051} (\bibinfo {year} {2021})},\ \Eprint
  {https://arxiv.org/abs/2103.06723} {arXiv:2103.06723 [astro-ph.IM]}
  \BibitemShut {NoStop}%
\bibitem [{\citenamefont {Golwala}(2000)}]{Golwala:2000zb}%
  \BibitemOpen
  \bibfield  {author} {\bibinfo {author} {\bibfnamefont {S.~R.}\ \bibnamefont
  {Golwala}},\ }\emph {\bibinfo {title} {{Exclusion limits on the WIMP nucleon
  elastic scattering cross-section from the Cryogenic Dark Matter Search}}},\
  \href {https://doi.org/10.2172/1421437} {Ph.D. thesis},\ \bibinfo  {school}
  {UC, Berkeley} (\bibinfo {year} {2000})\BibitemShut {NoStop}%
\bibitem [{\citenamefont {Gao}\ \emph {et~al.}(2007)\citenamefont {Gao},
  \citenamefont {Zmuidzinas}, \citenamefont {Mazin}, \citenamefont {LeDuc},\
  and\ \citenamefont {Day}}]{Gao:2007}%
  \BibitemOpen
  \bibfield  {author} {\bibinfo {author} {\bibfnamefont {J.}~\bibnamefont
  {Gao}}, \bibinfo {author} {\bibfnamefont {J.}~\bibnamefont {Zmuidzinas}},
  \bibinfo {author} {\bibfnamefont {B.~A.}\ \bibnamefont {Mazin}}, \bibinfo
  {author} {\bibfnamefont {H.~G.}\ \bibnamefont {LeDuc}},\ and\ \bibinfo
  {author} {\bibfnamefont {P.~K.}\ \bibnamefont {Day}},\ }\bibfield  {title}
  {\bibinfo {title} {{Noise properties of superconducting coplanar waveguide
  microwave resonators}},\ }\href {https://doi.org/10.1063/1.2711770}
  {\bibfield  {journal} {\bibinfo  {journal} {Applied Physics Letters}\
  }\textbf {\bibinfo {volume} {90}},\ \bibinfo {pages} {102507} (\bibinfo
  {year} {2007})}\BibitemShut {NoStop}%
\bibitem [{\citenamefont {Knapen}\ \emph {et~al.}(2021)\citenamefont {Knapen},
  \citenamefont {Kozaczuk},\ and\ \citenamefont {Lin}}]{Knapen:2021run}%
  \BibitemOpen
  \bibfield  {author} {\bibinfo {author} {\bibfnamefont {S.}~\bibnamefont
  {Knapen}}, \bibinfo {author} {\bibfnamefont {J.}~\bibnamefont {Kozaczuk}},\
  and\ \bibinfo {author} {\bibfnamefont {T.}~\bibnamefont {Lin}},\ }\bibfield
  {title} {\bibinfo {title} {{Dark matter-electron scattering in
  dielectrics}},\ }\href {https://doi.org/10.1103/PhysRevD.104.015031}
  {\bibfield  {journal} {\bibinfo  {journal} {Phys. Rev. D}\ }\textbf {\bibinfo
  {volume} {104}},\ \bibinfo {pages} {015031} (\bibinfo {year} {2021})},\
  \Eprint {https://arxiv.org/abs/2101.08275} {arXiv:2101.08275 [hep-ph]}
  \BibitemShut {NoStop}%
\bibitem [{\citenamefont {Boyd}\ \emph {et~al.}(2023)\citenamefont {Boyd},
  \citenamefont {Hochberg}, \citenamefont {Kahn}, \citenamefont {Kramer},
  \citenamefont {Kurinsky}, \citenamefont {Lehmann},\ and\ \citenamefont
  {Yu}}]{Boyd:2022tcn}%
  \BibitemOpen
  \bibfield  {author} {\bibinfo {author} {\bibfnamefont {C.}~\bibnamefont
  {Boyd}}, \bibinfo {author} {\bibfnamefont {Y.}~\bibnamefont {Hochberg}},
  \bibinfo {author} {\bibfnamefont {Y.}~\bibnamefont {Kahn}}, \bibinfo {author}
  {\bibfnamefont {E.~D.}\ \bibnamefont {Kramer}}, \bibinfo {author}
  {\bibfnamefont {N.}~\bibnamefont {Kurinsky}}, \bibinfo {author}
  {\bibfnamefont {B.~V.}\ \bibnamefont {Lehmann}},\ and\ \bibinfo {author}
  {\bibfnamefont {T.~C.}\ \bibnamefont {Yu}},\ }\bibfield  {title} {\bibinfo
  {title} {{Directional detection of dark matter with anisotropic response
  functions}},\ }\href {https://doi.org/10.1103/PhysRevD.108.015015} {\bibfield
   {journal} {\bibinfo  {journal} {Phys. Rev. D}\ }\textbf {\bibinfo {volume}
  {108}},\ \bibinfo {pages} {015015} (\bibinfo {year} {2023})},\ \Eprint
  {https://arxiv.org/abs/2212.04505} {arXiv:2212.04505 [hep-ph]} \BibitemShut
  {NoStop}%
\bibitem [{\citenamefont {DeRocco}\ \emph {et~al.}(2022)\citenamefont
  {DeRocco}, \citenamefont {Galanis},\ and\ \citenamefont
  {Lasenby}}]{DeRocco:2022rze}%
  \BibitemOpen
  \bibfield  {author} {\bibinfo {author} {\bibfnamefont {W.}~\bibnamefont
  {DeRocco}}, \bibinfo {author} {\bibfnamefont {M.}~\bibnamefont {Galanis}},\
  and\ \bibinfo {author} {\bibfnamefont {R.}~\bibnamefont {Lasenby}},\
  }\bibfield  {title} {\bibinfo {title} {{Dark matter scattering in
  astrophysical media: collective effects}},\ }\href
  {https://doi.org/10.1088/1475-7516/2022/05/015} {\bibfield  {journal}
  {\bibinfo  {journal} {JCAP}\ }\textbf {\bibinfo {volume} {05}}\bibfield
  {number} {\bibinfo  {number} { (05)},\ \bibinfo {pages} {015}},\ }\Eprint
  {https://arxiv.org/abs/2201.05167} {arXiv:2201.05167 [hep-ph]} \BibitemShut
  {NoStop}%
\bibitem [{\citenamefont {Feldman}\ and\ \citenamefont
  {Cousins}(1998)}]{Feldman:1997qc}%
  \BibitemOpen
  \bibfield  {author} {\bibinfo {author} {\bibfnamefont {G.~J.}\ \bibnamefont
  {Feldman}}\ and\ \bibinfo {author} {\bibfnamefont {R.~D.}\ \bibnamefont
  {Cousins}},\ }\bibfield  {title} {\bibinfo {title} {{A Unified approach to
  the classical statistical analysis of small signals}},\ }\href
  {https://doi.org/10.1103/PhysRevD.57.3873} {\bibfield  {journal} {\bibinfo
  {journal} {Phys. Rev. D}\ }\textbf {\bibinfo {volume} {57}},\ \bibinfo
  {pages} {3873} (\bibinfo {year} {1998})},\ \Eprint
  {https://arxiv.org/abs/physics/9711021} {arXiv:physics/9711021} \BibitemShut
  {NoStop}%
\end{thebibliography}%
\end{document}